\documentclass[11pt]{article}
\usepackage[margin=1in]{geometry}
\usepackage[small, labelfont={bf}, margin=1cm]{caption}
\usepackage[
  colorlinks,
  linkcolor = blue,
  citecolor = blue,
  urlcolor = blue]{hyperref}
\usepackage{amsmath,amssymb,amsthm}
\usepackage{thm-restate,colonequals,mathtools}
\usepackage{tikz}
\mathtoolsset{centercolon}

\usepackage{amssymb,amsmath,amsthm,enumerate}

\newtheorem{theorem}{Theorem}[section]
\newtheorem{lemma}[theorem]{Lemma}
\newtheorem{cor}[theorem]{Corollary}

\theoremstyle{definition}
\newtheorem*{definition}{Definition}
\newtheorem{defn}[theorem]{Definition}

\theoremstyle{conjecture}
\newtheorem{conjecture}[theorem]{Conjecture}

\newcommand{\xa}{x_{A}}
\newcommand{\xb}{x_{B}}
\newcommand{\ya}{y_{A}}
\newcommand{\yb}{y_{B}}

\newcommand{\qarrow}{\xrightarrow{q}}
\newcommand{\bartheta}{\bar{\vartheta}}
\DeclareMathOperator{\tr}{Tr}

\DeclareMathOperator{\vect}{vec}
\DeclareMathOperator{\rk}{rk}
\DeclareMathOperator{\og}{og}

\title{Quantum Homomorphisms}
\author{Laura Man\v{c}inska$^{1,2}$, David E. Roberson$^1$ \\ \normalsize
$^1$\textit{Department of Combinatorics \& Optimization and} \\ \normalsize
\textit{$^2$Institute for Quantum Computing} \\ \normalsize
\textit{University of Waterloo}}
\date{}

\begin{document}
\maketitle
\abstract{A homomorphism from a graph $X$ to a graph $Y$ is an adjacency preserving
mapping $f:V(X) \rightarrow V(Y)$. We consider a nonlocal game in which Alice and
Bob are trying to convince a verifier with certainty that a graph $X$
admits a homomorphism to $Y$. This is a generalization of the
well-studied graph coloring game. Via systematic study of quantum
homomorphisms we prove new results for graph coloring. Most
importantly, we show that the Lov\'{a}sz theta number of the complement lower bounds the
quantum chromatic number, which itself is not known to be computable.
We also show that some of our newly introduced graph parameters, namely quantum independence and clique numbers, can differ from their classical counterparts while others, namely quantum odd girth, cannot.
Finally, we show that quantum homomorphisms closely relate to
zero-error channel capacity. In particular, we use quantum
homomorphisms to construct graphs for which entanglement-assistance
increases their one-shot zero-error capacity.}

\tableofcontents

\section{Introduction}

The $c$-coloring game on a graph $X$ consists of two players, Alice and Bob, attempting to convince a referee that they have a $c$-coloring of $X$ \cite{Galliard02,Cleve04}. The game is played as follows: the referee sends each of the players a vertex of $X$, and each player responds with a color from $[c] = \{1,2,\ldots,c\}$. To win, the players must respond with the same color when they are sent the same vertex, and with different colors when they are sent adjacent vertices. The players are not allowed to communicate during the game but may agree on a strategy beforehand. It can be easily seen that  deterministic players with shared randomness (classical players) can win the $c$-coloring game on $X$ with certainty if and only if there exists a $c$-coloring of $X$. On the other hand, players allowed to make quantum measurements on some shared entangled state can sometimes win the $c$-coloring game on $X$ with certainty even when $X$ admits no $c$-coloring. Thus the quantum chromatic number, $\chi_q(X)$, is defined to be the smallest $c \in \mathbb{N}$ such that quantum players can win the $c$-coloring game on $X$ \cite{Avis}.

Quantum strategies for the coloring game and the quantum chromatic number have been well-studied \cite{Avis,qchrom,qchrom-aqis,SS12,MSS}. However, many questions remain unanswered. For example, it is not known whether $\chi_q(X)$ is computable, or whether there exists a family of graphs $X_n$ such that $\lim_{n \rightarrow \infty} \chi_q(X_n) < \infty$ but $\lim_{n \rightarrow \infty} \chi(X_n) = \infty$. Furthermore, there are few lower bounds known for quantum chromatic number. The authors of $\cite{qchrom}$ have shown that the orthogonal rank of a graph is a lower bound on a restricted version of quantum chromatic number, but the only general lower bound known is the clique number of a graph.

A \emph{homomorphism} from a graph $X$ to a graph $Y$ is a function, $\phi$, from the vertex set of $X$, denoted $V(X)$, to the vertex set of $Y$, denoted $V(Y)$, which preserves adjacency. More formally, $\phi: V(X) \rightarrow V(Y)$ is a homomorphism from $X$ to $Y$ if $\phi(x) \sim \phi(x')$ whenever $x \sim x'$, where ``$\sim$" denotes adjacency. We will write $X \rightarrow Y$ if there exists a homomorphism from $X$ to $Y$, and $X \not\rightarrow Y$ if not. It is straightforward to see that a homomorphism from a graph $X$ to the complete graph on $c$ vertices, denoted $K_c$, is equivalent to a $c$-coloring of $X$. Thus homomorphisms are a natural generalization of colorings. There is a well-developed and beautiful theory around graph homomorphisms \cite{Nesetril, Tardif}, and the study of them has given rise to original results on graph coloring.

Echoing the way in which homomorphisms generalize colorings, we have defined a homomorphism game which generalizes the coloring game. To play the $(X,Y)$-homomorphism game for graphs $X$ and $Y$, each player is sent a vertex of $X$, and must respond with a vertex of $Y$. In order to win, the players must respond with the same vertex of $Y$ when they are given the same vertex of $X$, and they must respond with adjacent vertices when given adjacent vertices. Similarly to the coloring game, classical players can win the $(X,Y)$-homomorphism game if and only if $X \rightarrow Y$. If quantum players can win the $(X,Y)$-homomorphism game, then we say that there is a \emph{quantum homomorphism} from $X$ to $Y$ and write $X \qarrow Y$.

Generalizing quantum colorings to quantum homomorphisms helps to place the idea into a broader and more natural mathematical context. Furthermore, a systematic study of quantum homomorphisms can potentially yield a better understanding of and new results concerning quantum colorings. Indeed, our Theorem~\ref{thetapreserved} can be applied to quantum colorings to obtain $\vartheta(\overline{X}) \le \chi_q(X)$, a previously unknown and efficiently computable lower bound. Here, $\overline{X}$ refers to the complement of $X$. Defining quantum homomorphisms also allows us to define quantum versions of many other graph parameters, including quantum independence number, which turns out to be intimately related to entanglement-assisted zero-error capacity. Lastly, quantum homomorphisms afford us many new examples of nonlocal games which potentially have perfect quantum strategies but no such classical strategy. We have already found an infinite family of such examples (see Section~\ref{alpha_q}).

\paragraph{Outline of the paper.} 
We start by introducing some notation as well as the mathematical formalism used to describe quantum states and measurements in Section~\ref{sec:Prelim}.
In Section~\ref{homogame} we introduce in detail the homomorphism game and the general forms of classical and quantum strategies. We also reformulate the question of the existence of a winning quantum strategy for a particular homomorphism in terms of the existence of a particular set of projectors satisfying certain orthogonality constraints. This is achieved via an argument previously used in \cite{qchrom} to obtain a similar reformulation for the coloring game.
We then use the reformulation to prove some basic properties of quantum homomorphisms. Additionally, we use this reformulation to show that there exists a quantum homomorphism from $X$ to $Y$ if and only if there exists a (classical) homomorphism from $X$ to an infinite graph which depends $Y$.

In Section~\ref{theta} we show that the order of the Lov\'{a}sz theta function, $\vartheta$, is respected by quantum homomorphisms, i.e.~if $X \qarrow Y$, then $\vartheta(\overline{X}) \le \vartheta(\overline{Y})$. This is followed by section~\ref{parameters} in which we define and investigate quantum analogs of some graph parameters that can be defined using homomorphisms. We start with quantum chromatic number, which has been previously defined. We use the main result of Section~\ref{theta} to show that the quantum chromatic number of $X$ is lower bounded by $\vartheta(\overline{X})$. Using this bound we are able to show that the quantum colorings given in \cite{Avis} are optimal. Moving on to quantum independence number, we prove a quantum analog of a classical result relating the existence of a homomorphism from $X$ to $Y$ to the existence of an independent set of size $|V(X)|$ in a graph depending on $X$ and $Y$. We then use this to construct graphs which have large separation between classical and quantum independence number. We again use the result from Section~\ref{theta} to show that quantum independence number is upper bounded by Lov\'{a}sz theta number. The last parameter we consider is quantum odd girth. In contrast to quantum chromatic and independence numbers, we show that there are no separations between odd girth and quantum odd girth. To prove this we first prove a quantum analog of a well-known result relating homomorphisms and walks in graphs.

In Section~\ref{capacity} we show that the entanglement-assisted one-shot zero-error capacity of a graph is lower bounded by the quantum independence number. Furthermore, we show that if the entanglement-assisted one-shot zero-error capacity of a graph can be achieved using only projective measurements on a maximally entangled state, then it is equal to the quantum independence number. We also show that the order of entanglement-assisted one-shot zero-error capacity is respected by quantum homomorphisms, i.e.,~that $c_0^*(\overline{X}) \le c_0^*(\overline{Y})$ whenever $X \qarrow Y$.

In Section~\ref{rank} we introduce a new graph parameter which we refer to as \emph{projective rank} and denote $\xi_f$, that is a generalization of both orthogonal rank and fractional chromatic number. We show that this parameter lies between Lov\'{a}sz theta of the complement and quantum chromatic number. We also show that the order of projective rank is respected by quantum homomorphisms, i.e.,~if $X \qarrow Y$, then $\xi_f(X) \le \xi_f(Y)$.

\subsection{Preliminaries}
\label{sec:Prelim}

In this paper graphs will be undirected and, unless explicitly defined otherwise, finite. A graph is said to be vertex/edge transitive if for any two vertices/edges, there exists an automorphism of the graph that maps one to the other. We will refer to orthogonal projectors (matrices which are both idempotent and Hermitian) simply as projectors. We say that two positive semidefinite matrices $E$ and $F$ of the same dimensions, are orthogonal if $EF=0$ or equivalently if $\tr(EF)=0$. Note that this also implies that the column spaces of $E$ and $F$ are orthogonal.
We use $\mathbb{M}_{\mathbb{F}}(m,n)$ to denote the space of $m\times n$ matrices over the field $\mathbb{F}$. For a matrix $M\in\mathbb{M}_{\mathbb{C}}(m,n)$, we write $\overline{M}$ to denote the $m\times n$ matrix obtained by taking the complex conjugate of each of the entries of $M$. Finally, for a natural number $n\ge 1$ we use $[n]$ to denote the set $\{1,\dotsc,n\}$.

\paragraph{Quantum states and measurements.}
We now briefly introduce the mathematical formalism used to describe quantum states and measurements. For an in-depth explanation we refer the reader to~\cite{NC}.

To any quantum system S, we associate a complex vector space $\mathbb{C}^d$ equipped with the usual inner product. A state of the system S can be specified by a unit vector $\psi\in\mathbb{C}^d$ or alternatively using a rank-one projection $\psi\psi^*$. These are the most basic type of quantum states and they are known as  pure states.
Now suppose that the quantum system S is prepared in state $\psi\in\mathbb{C}^d$ with probability $p$ and in state $\phi\in\mathbb{C}^d$ with probability  $1-p$. If we do not know which case has happened, we describe the state of S using the positive semidefinite matrix $p \psi\psi^* + (1-p)\phi\phi^*$. Such states are known as mixed states. More formally, a mixed state of system S is a positive semidefinite matrix $\rho\in\mathbb{M}_{\mathbb{C}}(d,d)$ with trace equal to one. We can extract classical information from a quantum system by measuring it. In this paper we describe a measurement using a positive-operator valued measure (POVM) and we thus use the terms ``measurement'' and ``POVM'' interchangeably. A POVM $\mathcal{E}$ is specified by a set $\{E_1,\ldots,E_n\}\subset\mathbb{M}_{\mathbb{C}}(d,d)$ of positive semidefinite operators such that $\sum_{i \in [n]} E_i = I_d$, where $I_d$ is the $d\times d$ identity matrix. A POVM, or measurement, is projective if each $E_i$ is a projector. Note that we can only measure a quantum system S using a measurement $\mathcal{E}$ if the dimension of the vector space associated to S matches the size of the matrices specifying the measurement $\mathcal{E}$. 
According to the axioms of quantum mechanics, if we measure a quantum system S in a mixed state $\rho$ using measurement $\mathcal{E}$, then we obtain outcome $i\in [n]$ with probability equal to $p_i = \tr(\rho E_i)$. In the case when the state $\rho$ is pure, i.e., $\rho = \psi\psi^*$, we can also express the probability $p_i$ as~$\psi^* E_i \psi$. 

Consider two quantum systems $\text{A}$ and $\text{B}$ with associated spaces $\mathbb{C}^{d_A}$ and $\mathbb{C}^{d_B}$ respectively. The vector space we associate to the combined quantum system $(\text{A},\text{B})$ is $\mathbb{C}^{d_A}\otimes \mathbb{C}^{d_B}$. Furthermore, if $\text{A}$ is in a state $\psi_1\in\mathbb{C}^{d_A}$ and $\text{B}$ is in state $\psi_2\in\mathbb{C}^{d_B}$, then their joint state is described using the tensor product $\psi_1\otimes\psi_2$. However, not all pure states of $(\text{A},\text{B})$ are tensor products. In fact, one of the key features of quantum mechanics, the phenomenon of entanglement, occurs when the state of the constituent quantum systems cannot be described independently. More precisely, a state $\psi\in\mathbb{C}^{d_A}\otimes \mathbb{C}^{d_B}$ is said to be entangled, if there does not exist a pair of states $\psi_1\in\mathbb{C}^{d_A}, \psi_2\in\mathbb{C}^{d_B}$ such that $\psi = \psi_1\otimes\psi_2$. We say that a quantum state $\psi \in \mathbb{C}^d \otimes \mathbb{C}^d$ is \emph{maximally entangled} if $\psi = \frac{1}{\sqrt{d}}\sum_{i = 1}^d u_i \otimes v_i$, where $\{u_1,\ldots,u_d\}$ and $\{v_1,\ldots,v_d\}$ are some orthonormal bases for $\mathbb{C}^d$. The \emph{canonical maximally entangled state} in $\mathbb{C}^d \otimes \mathbb{C}^d$ is 
\begin{equation}
  \Phi := \frac{1}{\sqrt{d}}\sum_{i=1}^d e_i \otimes e_i,
\end{equation}
where $e_i$ is the $i^\text{th}$ standard basis vector. 

Finally, let us discuss partial measurements of a composite quantum system $(\text{A},\text{B})$. If a POVM $\mathcal{E} = \{E_1,\dotsc,E_n\}$ can be used to measure system A then it can also be used to measure the A part of the composite system $(\text{A},\text{B})$. Applying $\mathcal{E}$ to a state $\psi\in\mathbb{C}^{d_A}\otimes \mathbb{C}^{d_B}$ of $(\text{A},\text{B})$ will yield outcome $i\in[n]$ with probability
\begin{equation}
  p_i = \psi^* (E_i \otimes I) \psi
\end{equation} 
leaving system B in state $\rho_i = \tfrac{1}{p_i} \tr_A((E_i \otimes I) \psi\psi^*)$. Here, $\tr_A: \mathbb{M}_{\mathbb{C}}(d_Ad_B,d_Ad_B) \to \mathbb{M}_{\mathbb{C}}(d_B,d_B)$ is the map defined by $\tr_A(M\otimes N) =\tr(M)N$ and extended linearity; this map is known as partial trace. We sometimes refer to the $\rho_i$ as the residual states. Note that there is no need to define residual states, when $p_i=0$ as this corresponds to a zero-probability event. To treat both the $p_i=0$ and the $p_i\ne 0$ case in a unified manner, it is sometimes convenient to consider the unnormalized residual states $\rho'_i := \tr_A((E_i \otimes I) \psi\psi^*)$. Note that the states $\rho'_i$ are well-defined and equal to the zero matrix for $p_i=0$. We stress that unnormalized states are used for convenience only and all quantum states are by definition normalized, i.e., of unit norm.

\section{Homomorphism Game}\label{homogame}

For a graph $X$ and positive integer $c$, the $(X,c)$-coloring game is played as follows: a verifier sends Alice and Bob vertices $\xa$ and $\xb$ of $X$ respectively, and they respond with colors \mbox{$\ya,\yb \in [c]$} accordingly. The players may decide on a strategy beforehand, but are not allowed to communicate after receiving input from the verifier. To win, the following conditions need to be satisfied:
\begin{align}
&\text{if } \xa = \xb, \text{ then } \ya = \yb; \\
&\text{if } \xa \sim \xb, \text{ then } \ya \ne \yb.
\end{align}
The first condition is usually referred to as the consistency condition, while the second  condition corresponds to the requirements of a proper coloring of a graph. The game is only played for one round, i.e.~one pair of inputs and one pair of outputs are involved. It is of particular interest when quantum players can win the $c$-coloring game for some graph $X$ for which classical players cannot. As mentioned above, the quantum chromatic number is the smallest $c \in \mathbb{N}$ such that quantum players can win the $(X,c)$-coloring game.

For an ordered pair of graphs $(X,Y)$, the $(X,Y)$-homomorphism game, consists of a verifier sending Alice and Bob vertices $\xa$ and $\xb$ of $X$ respectively, and Alice and Bob responding with vertices $\ya$ and $\yb$ of $Y$ accordingly. To win the $(X,Y)$-homomorphism game, the following conditions need to be satisfied:
\begin{align}
&\text{if } \xa = \xb, \text{ then } \ya = \yb; \\
&\text{if } \xa \sim \xb, \text{ then } \ya \sim \yb.
\end{align}
Like above, the first condition is the consistency condition and the second corresponds to the adjacency-preserving property of homomorphisms. When $Y = K_c$, this game reduces to the $(X,c)$-coloring game since inequality is adjacency in the complete graph. When we say that two players can ``win'' a game, we mean that they can win with probability 1.

\subsection{The Classical Strategy}

Any deterministic strategy for the $(X,Y)$-homomorphism game can be specified by some functions $f_A, f_B : V(X) \rightarrow V(Y)$, where Alice responds with $f_A(x)$ whenever she receives $x$ as an input, and $f_B$ is defined similarly. Due to the consistency condition, we must have that $f_A = f_B$ for a winning strategy. Furthermore, it must be the case that $f_A$ is in fact a homomorphism from $X$ to $Y$. Conversely, if $\phi : X \rightarrow Y$ is a homomorphism, then each player responding with $\phi(x)$ upon receiving $x$ is a winning deterministic strategy for the $(X,Y)$-homomorphism game.

Classical players are in general not restricted to deterministic strategies, they are allowed probabilistic strategies which depend on shared randomness. However, in this case their strategy can be viewed as a probability distribution over deterministic strategies. Since they must win with certainty, each deterministic strategy which occurs with nonzero probability must be a winning deterministic strategy. Therefore, by the above there must still exist a homomorphism from $X$ to $Y$ in this case. Thus classical players can win the $(X,Y)$-homomorphism game if and only if $X \rightarrow Y$.

\subsection{The Quantum Strategy}

For quantum Alice and Bob, the most general strategy for playing the $(X,Y)$-homomorphism game is as follows: upon receiving input $x \in V(X)$, Alice performs POVM $\mathcal{E}_x = \{E_{xy}\}_{y \in V(Y)}$ on her part of a shared state $\psi \in \mathbb{C}^{d_A} \otimes \mathbb{C}^{d_B}$ and obtains some outcome $y \in V(Y)$, which she sends to the verifier as her answer. Bob acts similarly, except that he uses POVMs $\mathcal{F}_x = \{F_{xy}\}_{y \in V(Y)}$ for $x \in V(X)$. The probability that Alice outputs $y \in V(Y)$ and Bob outputs $y' \in V(Y)$ upon receiving inputs $x,x' \in V(X)$ respectively is given by $\psi^*(E_{xy} \otimes F_{x'y'})\psi$.
Therefore, to win we need that
\begin{align*}
\psi^* \left(E_{xy} \otimes F_{x'y'}\right) \psi &= 0 \text{  for } y \ne y'; \\
\psi^* \left(E_{xy} \otimes F_{x'y'}\right) \psi &= 0 \text{  for } x \sim x' \text{ and } y \not\sim y',
\end{align*}
where $y \not\sim y'$ includes the case where $y = y'$. Since quantum players can always perform at least as well as classical ones, we have that $X \rightarrow Y \Rightarrow X \qarrow Y$. 

The following theorem was proven by Cameron et. al. in \cite{qchrom} for the coloring game, and the same proof works more generally for the homomorphism game.

\begin{restatable}{theorem}{NormalForm}\label{thm:normal}
If the $(X,Y)$-homomorphism game can be won by a quantum strategy, then it can be won by a quantum strategy such that
\begin{enumerate}[\indent a.]
\item the matrices $E_{xy}$ and $F_{xy}$ are $d \times d$ projectors for some $d \in \mathbb{N}$;
\item $E_{xy} = \overline{F}_{xy}$ for all $x \in V(X)$, $y \in V(Y)$;
\item $\psi = \frac{1}{\sqrt{d}} \sum_{i = 1}^d e_i \otimes e_i$.
\end{enumerate}
Furthermore, if $Y$ is vertex transitive, then we may assume that:
\begin{enumerate}[\indent a.]
\item[d.] The projectors $E_{xy}$ all have the same rank.
\end{enumerate}
\end{restatable}

Further following the argument of \cite{qchrom} we obtain:
\begin{cor}\label{cor:reformulation}
There exists a quantum homomorphism from $X$ to $Y$ if and only if there exist projectors $E_{xy}$ for $x \in V(X), y \in V(Y)$ such that
\begin{align}
\label{cond:ident} \sum_{y \in V(Y)} E_{xy} &= I \ \forall x \in V(X);\\
\label{cond:orthog} E_{xy}E_{x'y'} &= 0 \text{ if } (x = x' \ \& \ y \ne y') \text{ or } (x \sim x' \ \& \ y \not\sim y').
\end{align}
\end{cor}

We say that ``$E_{xy}$ are projectors that give a quantum homomorphism from $X$ to $Y$" if the matrices $E_{xy}$ satisfy the conditions of Corrolary~\ref{cor:reformulation}.

The following lemma shows that we can further restrict to using only real projectors. We will need this fact later when proving results about Lov\'{a}sz theta.

\begin{lemma}\label{real}
There exists a quantum homomorphism from $X$ to $Y$ if and only if there exist real projectors $E_{xy}$ for all $x \in V(X), y \in V(Y)$ satisfying (\ref{cond:ident}) and (\ref{cond:orthog}) from Corollary \ref{cor:reformulation}.
\end{lemma}
\proof

The `if' direction is trivial by Corollary \ref{cor:reformulation}. To prove the other direction, suppose that $X \qarrow Y$. By Corollary \ref{cor:reformulation} there exist (possibly complex) projectors $E_{xy} \in \mathbb{M}_\mathbb{C}(d,d)$ for $x \in V(X), y \in V(Y)$ satisfying (\ref{cond:ident}) and (\ref{cond:orthog}) above. Let $R : \mathbb{M}_\mathbb{C}(d,d) \rightarrow \mathbb{M}_\mathbb{R}(2d,2d)$ be the map defined by
\[R(A) = \left(\begin{array}{cc}
\Re(A) & \Im(A) \\
-\Im(A) & \Re(A)
\end{array}\right)\]
where $\Re(A)$ and $\Im(A)$ are the real and imaginary parts of $A$ respectively. It is routine to check that $R(A + B) = R(A) + R(B)$, $R(AB) = R(A)R(B)$, and that $R$ takes Hermitian matrices to symmetric matrices. Therefore, for all $x \in V(X)$ and $y \in V(Y)$, the matrix $R(E_{xy})$ is a real projector that satisfies (\ref{cond:ident}) and (\ref{cond:orthog}).\qed

Recall from Theorem \ref{thm:normal} that if $Y$ is vertex transitive we can assume that the projectors used in a winning quantum strategy all have the same rank. This gives us the following corollary.

\begin{cor}\label{reformvtxtrans}
If $Y$ is vertex transitive, then $X \qarrow Y$ if and only if there exist rank $r$ projectors $E_{xy} \in \mathbb{M}_\mathbb{C}(d,d)$ for $x \in V(X), y \in V(Y)$ for some $r,d \in \mathbb{N}$ such that
\begin{align}
&d = r|V(Y)|; \\
&E_{xy}E_{x'y'} = 0 \text{ if } (x = x' \ \& \ y \ne y') \text{ or } (x \sim x' \ \& \ y \not\sim y').
\end{align}
\end{cor}
\proof
Suppose that such projectors exist. Since the sum of mutually orthogonal projectors is equal to identity if and only if the sum of their ranks is equal to the dimension in which they live,  we have that
\[\sum_{y \in V(Y)} E_{xy} = I \ \forall x \in V(X).\]
Therefore by Corollary \ref{cor:reformulation} we have that $X \qarrow Y$.

If $X \qarrow Y$, then by Corollary \ref{cor:reformulation} we have that there exist projectors $E_{xy} \in \mathbb{M}_\mathbb{C}(d,d)$ for $x \in V(X), y \in V(Y)$ for some $d \in \mathbb{N}$ that satisfy (\ref{cond:ident}) and (\ref{cond:orthog}). Furthermore, since $Y$ is vertex transitive, we can assume that the projectors $E_{xy}$ all have the same rank, $r$, for some $r \in \mathbb{N}$. Condition (\ref{cond:ident}) then implies that $d = r|V(Y)|$.\qed

\subsection{Basic Properties of Quantum Homomorphisms}

In this section we discuss some basic properties of quantum homomorphisms which we need later. First, we show that, like classical homomorphisms, quantum homomorphisms are transitive.

\begin{lemma}\label{lem:trans}
The relation $\qarrow$ is transitive: if $X \qarrow Y$ and $Y \qarrow Z$, then $X \qarrow Z$.
\end{lemma}

\proof
Assume that $X \qarrow Y$ and $Y \qarrow Z$. Consider the $(X,Z)$-homomorphism game; we now describe a strategy to win this game which relies only on the fact that Alice and Bob can win both the $(X,Y)$- and $(Y,Z)$-homomorphism games. Alice is given $\xa \in V(X)$ and Bob is given $\xb \in V(X)$. Alice and Bob first act as if they are playing the $(X,Y)$-homomorphism game to obtain outputs $\ya,\yb \in V(Y)$ respectively. They then act as if they are playing the $(Y,Z)$-homomorphism game with inputs $\ya$ and $\yb$ to obtain outputs $z_A,z_B \in V(Z)$ respectively. If $\xa = \xb$, then $\ya = \yb$ and then $z_A = z_B$ since Alice and Bob win the $(X,Y)$-homomorphism game and $(Y,Z)$-homomorphism game. Similarly, if $\xa \sim \xb$, then $\ya \sim \yb$ and so $z_A \sim z_B$. Therefore, Alice and Bob can win the $(X,Z)$-homomorphism game, i.e. $X \qarrow Z$.\qed

The above proof does not use that fact that the players are using a particular type of strategy, therefore this also proves that classical homomorphisms are transitive, though that is well-known and easy to see. In the quantum case, the method of ``composition" given in the proof corresponds to the following: if $\{E_{xy}: x \in V(X), y \in V(Y)\}$ are the projectors which give a quantum homomorphism from $X$ to $Y$, and $\{E'_{yz}: y \in V(Y), z \in V(Z)\}$ are the projectors which give a quantum homomorphism from $Y$ to $Z$, then the following projectors give a quantum homomorphism from $X$ to $Z$:
\[E''_{xz} = \sum_{y \in V(Y)} E_{xy}\otimes E'_{yz} \text{ for } x \in V(X), z \in V(Z).\]

The next result was given in different terms in \cite{qchrom}. It shows that when restricted to complete graphs, quantum homomorphisms behave the same as homomorphisms.

\begin{lemma}\label{complete}
$K_m \qarrow K_n$ if and only if $m \le n$.
\end{lemma}
\proof
If $m \le n$, then $K_m \rightarrow K_n$ and thus $K_m \qarrow K_n$.

Now suppose that $K_m \qarrow K_n$. Let $E_{xy}$ for $x \in V(K_m), y \in V(K_n)$ be the projectors satisfying the conditions of Corollary \ref{reformvtxtrans}. The projectors $E_{xy}$ have rank some $r \in \mathbb{N}$ and live in $\mathbb{M}_\mathbb{C}(rn,rn)$. For fixed $y \in V(K_n)$ the projectors $\{E_{xy}\}_{x \in V(K_m)}$ are pairwise orthogonal. Therefore,
\[\rk\left(\sum_{x \in V(K_m)} E_{xy}\right) = \sum_{x \in V(K_m)} \rk\left(E_{xy}\right) = rm\]
and thus $rm \le rn$ which of course implies that $m \le n$.\qed

\begin{lemma}
Suppose that $X$ is a connected graph and $Y$ is a graph with connected components $Y_1,\ldots,Y_n$. If $X \qarrow Y$, then there exists $k \in [n]$ such that $X \qarrow Y_k$.
\end{lemma}
\proof
If $X$ is a single vertex then we are done, otherwise $X$ has an edge. Suppose that $X \qarrow Y$ and that $E_{xy}$ for $x \in V(X), y \in V(Y)$ are projectors satisfying the conditions of Corollary \ref{cor:reformulation}. For each $i \in [n]$ and $x \in V(X)$ define $E_{xi}$ as follows:
\[E_{xi} = \sum_{y \in V(Y_i)} E_{xy}.\]
Since $E_{xy}E_{xy'} = 0$ for $y \ne y'$, we have that the matrices $E_{xi}$ are projectors. Furthermore, since the $V(Y_i)$ partition $V(Y)$, we have that $E_{xi}E_{xj} = 0$ for $i \ne j$. Now suppose that $x',x'' \in V(X)$ are adjacent. Since $y \not\sim y'$ for $y \in V(Y_i), y' \in V(Y_j), i \ne j$ we have that
\[E_{x'i}E_{x''j} = 0 \text{ for all } i \ne j.\]
This implies that $E_{x'i}\left(I - E_{x''i}\right) = 0$ and $\left(I - E_{x'i}\right)E_{x''i} = 0$ for all $i \in [n]$, as $\sum_{i\in[n]} E_{x'i}= \sum_{j\in[n]} E_{x''j}=I$. Therefore, $E_{x'i} = E_{x'i}E_{x''i} = E_{x''i}$.  Since $X$ is connected, we further obtain that for each $i\in[n]$ there exists a projector $E_i$ such that $E_{xi} = E_i$ for all $x\in V(X)$.

Since $\sum_{i \in [n]} E_{i} = I$, we can fix a $k \in [n]$ such that $E_{k} \ne 0$. Then for all $x\in V(X), y\in V(Y_k)$ we have that
\begin{align*}
&\sum_{y \in V(Y_k)} E_{xy} = E_k \\
&E_{xy}E_{x'y'} = 0 \text{ if } (x = x' \ \& \ y \ne y') \text{ or } (x \sim x' \ \& \ y \not\sim y').
\end{align*}
Hence, when restricted to the image of $E_k$, the projectors $E_{xy}$ for $x\in V(X),y\in V(Y_k)$ satisfy the conditions of Corollary \ref{cor:reformulation} and $X \qarrow Y_k$ as desired.\qed

\subsection{Reformulating Quantum Homomorphisms in Terms of Homomorphisms}

Using Corollary~\ref{cor:reformulation}, we can rephrase the question of the  existence of a quantum homomorphism from $X$ to $Y$ as a question of the existence of a homomorphism from $X$ to a graph, $\mathfrak{M}(Y)$, which depends only on $Y$. In theory, this may offer some assistance in the study of quantum homomorphisms, since it allows one to apply results concerning homomorphisms, which have been extensively studied. In practice it may not be of much use because the graphs $\mathfrak{M}(Y)$ seem to be difficult to analyse, and in particular contain an uncountable number of vertices.

\begin{definition}
For a finite graph $Y$ and $d \in \mathbb{N}$, define $M(Y,d)$ to be the graph whose vertices are functions $p:V(Y) \rightarrow \mathbb{M}_\mathbb{C}(d,d)$ such that $p(y)$ is a projector for all $y \in V(Y)$, and $\sum_{y \in V(Y)} p(y) = I$. Two such functions, $p$ and $p'$, are adjacent if $p(y)p'(y') = 0$ for all $y \not\sim y'$. Further, define $\mathfrak{M}(Y)$ to be the disjoint union of $M(Y,d)$ over all $d \in \mathbb{N}$.
\end{definition}

Note that the vertices of $M(Y,d)$ are exactly projective measurements in dimension $d$ whose outcomes are indexed by the vertices of $Y$. We have the following:

\begin{theorem}\label{qhom2homd}
If $X$ and $Y$ are graphs, then $X \qarrow Y$ if and only if $X \rightarrow M(Y,d)$ for some $d \in \mathbb{N}$.
\end{theorem}

\proof
Suppose that $\phi: X \rightarrow M(Y,d)$ for some $d \in \mathbb{N}$. For all $x \in V(X)$ and $y \in V(Y)$, let $E_{xy} = \phi(x)(y)$. It is straightforward to check that the projectors $E_{xy}$ satisfy the conditions of Corollary~\ref{cor:reformulation} and thus $X \qarrow Y$.

Conversely, if $X \qarrow Y$, then there exist projectors $E_{xy}$ satisfying the conditions of Corollary~\ref{cor:reformulation}. For each $x \in V(X)$, define $p_x : V(Y) \rightarrow \mathbb{M}_\mathbb{C}(d,d)$ as $p_x(y) = E_{xy}$. The function $\phi : V(X) \rightarrow V(M(Y,d))$ defined by $\phi(x) = p_x$ is clearly a homomorphism from $X$ to $M(Y,d)$.
\qed

If we really want to only consider a single target graph instead of a family, then we can use $\mathfrak{M}(Y)$:

\begin{theorem}\label{qhom2hom}
If $X$ and $Y$ are graphs, then $X \qarrow Y$ if and only if $X \rightarrow \mathfrak{M}(Y)$.
\end{theorem}

\proof
If $X \qarrow Y$ then by Theorem~\ref{qhom2homd} there exists $d \in \mathbb{N}$ such that $X \rightarrow M(Y,d)$. Since $M(Y,d)$ is a subgraph of $\mathfrak{M}(Y)$, we have that $X \rightarrow \mathfrak{M}(Y)$.

Now suppose that $\phi: X \rightarrow \mathfrak{M}(Y)$. If $X'$ is a component of $X$, then the image of $X'$ under $\phi$ is connected and therefore must be contained in $M(Y,d)$ for some $d \in \mathbb{N}$. Therefore $X' \qarrow Y$ by Theorem~\ref{qhom2homd}, and similarly all components of $X$ have a quantum homomorphism from $X$ to $Y$. Now by considering the $(X,Y)$-homomorphism game, it is easy to see that $X$ must have a quantum homomorphism to $Y$: each player plays according to the strategy required for the component of $X$ from which their input vertex comes.\qed

\subsection{Non-signalling Strategies}\label{nonsignal}

We here consider the case in which Alice and Bob have access to more general correlations than those produced by shared entanglement. Mathematically, a correlation can be described as a joint conditional probability distribution $P(Y_A, Y_B | X_A, X_B)$, where $X_A$ and $X_B$ are random variables representing Alice and Bob's respective inputs to the correlation, and $Y_A$ and $Y_B$ their respective outputs. For instance, when Alice and Bob employ a quantum strategy for some game, then the correlation produced by their strategy is given by
\[P(Y_A = y, Y_B = y' | X_A = x, X_B = x') = \psi^* \left(E_{xy} \otimes F_{x'y'}\right)\psi^*,\]
where $E_{xy}$ and $F_{x'y'}$ are Alice and Bob's POVM elements corresponding to respective inputs $x$ and $x'$ and outputs $y$ and $y'$, and $\psi$ is their shared state. The \emph{set of quantum correlations} (for some fixed input and output sets) is defined to be the set of all correlations that can be produced in this manner. An important property of quantum correlations is that they are non-signalling, i.e.,~Alice cannot obtain any information concerning Bob's input based on her input and output, and vice versa. More formally, we have the following definition:

\begin{definition}
A correlation $P$ is \emph{non-signalling} (with respect to the partition $\{(X_A,Y_A),(X_B,Y_B)\}$) if
\begin{align*}
P(Y_A = y | X_A = x, X_B = x') &:= \sum_{y'} P(Y_A = y, Y_B = y' | X_A = x, X_B = x') \\
P(Y_B = y | X_A = x', X_B = x) &:= \sum_{y'} P(Y_A = y', Y_B = y | X_A = x', X_B = x)
\end{align*}
are both independent of $x'$.
\end{definition}
The widely accepted physical principle which motivates this definition is that information cannot be transmitted faster than light. Thus, if Alice and Bob are sufficiently far apart and must produce their outputs within a short time span, then they can only produce non-signalling correlations.

It is sometimes useful to consider what parties are able to achieve using arbitrary non-signalling correlations. This relaxation lets us place bounds on what is achievable using quantum correlations, and often allows for easier analysis. Note, however, that arbitrary non-signalling correlations may not actually be producible by any physical means, but of course the same could have been said of quantum correlations just over 100 years ago. So when we say that Alice and Bob have access to arbitrary non-signalling correlations, we think of each of them as having a black box into which takes their respective inputs and provides them with an output. The only requirement we place on these (potentially physically impossible) boxes is that the correlation they produce is non-signalling.

It turns out that players with access to such arbitrary non-signalling correlations can play the homomorphism game just as well as if they were able to communicate during the game. To see this note that for any graph $X$, the $(X,K_2)$-homomorphism game can be won using the following non-signalling correlation:
\begin{displaymath}
   P(Y_A = \ya, Y_B = \yb | X_A = \xa, X_B = \xb) = \left\{
     \begin{array}{ll}
       \frac{1}{2} & \text{if } \delta(\xa,\xb)  = \delta(\ya,\yb) \\
       0 &  \text{o.w.}
     \end{array}
   \right.
\end{displaymath}
Here $X_A, X_B$ take values from $V(X)$ and $Y_A, Y_B$ take values from $V(K_2)$, and $\delta$ is the Kronecker delta function. Note that there is no way to win the $(X,K_1)$-homomorphism game if $X$ is not empty, since if the players are given adjacent vertices they will always lose.

\section{Lov\'{a}sz $\vartheta$}\label{theta}

The Lov\'{a}sz theta number of a graph $X$, denoted $\vartheta(X)$, was introduced in \cite{lovasz} as an upper bound on the Shannon capacity $\Theta(X)$. It was later shown to upper bound the entanglement-assisted Shannon capacity $\Theta^*(X)$ as well \cite{Winter,Beigi}. Since $\vartheta$ admits a semidefinite program formulation, it can be approximated efficiently and is thus of great practical use. In this paper we will be mostly interested in Lov\'{a}sz theta of the complement of a graph, denoted $\bartheta(X)$, as this is preserved by homomorphisms. There are various definitions for $\vartheta$, and hence $\bartheta$, but the one below from~\cite{chivec} suits us since it directly defines $\bartheta$ in terms of homomorphisms.

\begin{defn}\label{LVhomo}
For $m \in \mathbb{N}$ and $\alpha < 0$, let $S(m,\alpha)$ be the infinite graph whose vertices are unit vectors in $\mathbb{R}^m$ and two vectors $u$ and $v$ are adjacent if $u^T v = \alpha$. Then
\[\bartheta(X) = \inf \left\{1 - \frac{1}{\alpha} : X \rightarrow S(m,\alpha), \ \alpha < 0 \right\}.\]
In fact, the infimum in the definition above can be replaced by a minimum unless $X$ is an empty graph, in which case $\bartheta(X) = 1$. Also, note that $m$ is not restrained, but we can assume that $m = |V(X)|$ if we like, since any set of $|V(X)|$ vectors lives in a space of dimension at most $|V(X)|$. 
\end{defn}

From this definition it is clear that if $X \rightarrow Y$, then $\bartheta(X) \le \bartheta(Y)$. Surprisingly, the same is true for quantum homomorphisms.

\begin{theorem}\label{thetapreserved}
If $X \qarrow Y$, then $\bartheta(X) \le \bartheta(Y)$.
\end{theorem}
\proof
Suppose that $X \qarrow Y$ and $Y \rightarrow S(m,\alpha)$ for some $m \in \mathbb{N}$ and $\alpha < 0$. To prove the theorem it suffices to show that $X \rightarrow S(m',\alpha)$ for some $m' \in \mathbb{N}$. Since $Y \rightarrow S(m,\alpha)$, there exist unit vectors $v_y \in \mathbb{R}^m$ for $y \in V(Y)$ such that $v_y^T v_{y'} = \alpha$ whenever $y \sim y'$. Since $X \qarrow Y$, Lemma~\ref{real} implies that there exist \emph{real} projectors $E_{xy}$ in some dimension $d$ that give a quantum homomorphism from $X$ to $Y$. Define vectors $u_x$ for $x \in V(X)$ by
\[u_x = \frac{1}{\sqrt{d}}\sum_{y \in V(Y)} v_y \otimes \vect(E_{xy})\]
where $\vect(M)$ is the vector obtained by stacking the rows of $M$ (as column vectors) on top of each other. Then we have
\begin{align*}
u_x^T u_{x'} &= \frac{1}{d}\left(\sum_{y \in V(Y)} v_y \otimes \vect(E_{xy})\right)^T \left(\sum_{y' \in V(Y)} v_{y'} \otimes \vect(E_{x'y'})\right) \\
&= \frac{1}{d}\sum_{y,y' \in V(Y)} v_y^T v_{y'} \tr(E_{xy}E_{x'y'}).
\end{align*}
Since $\tr(E_{xy}E_{xy'}) = 0$ for all $y \ne y'$, we further get
\begin{align*}
u_x^T u_x &= \frac{1}{d}\sum_{y \in V(Y)} v_y^T v_y \tr(E_{xy}E_{xy}) \\
&= \frac{1}{d}\tr\left(\sum_{y \in V(Y)} E_{xy}\right) \\
&= \frac{1}{d}\tr(I) = 1.
\end{align*}
So the vectors $u_x$ have unit norm and now we just need to check that $u_x^T u_{x'} = \alpha$ whenever $x \sim x'$. In this case, we have that $\tr(E_{xy}E_{x'y'}) = 0$ whenever $y \not\sim y'$ and so
\begin{equation}\label{eq:theta}
\begin{split}
u_x^T u_{x'} &= \frac{1}{d} \sum_{y \sim y'} v_y^T v_{y'} \tr(E_{xy}E_{x'y'}) = \frac{\alpha}{d} \sum_{y \sim y'} \tr(E_{xy}E_{x'y'}) \\
&= \frac{\alpha}{d} \sum_{y,y' \in V(Y)} \tr(E_{xy}E_{x'y'}) = \frac{\alpha}{d} \tr\left(\left(\sum_{y \in V(Y)}E_{xy}\right)\left(\sum_{y' \in V(Y)}E_{x'y'}\right)\right) \\
&= \alpha.
\end{split}
\end{equation}
Hence $X \rightarrow S(md^2,\alpha)$ and we are done.
\qed

As we will see in the next section, this theorem can be used to bound quantum parameters defined in terms of homomorphisms, such as quantum chromatic number. In general, no algorithm is known for deciding whether $X \qarrow Y$. However, if $\bartheta(X) > \bartheta(Y)$, the above theorem allows us to conclude that $X$ does not admit a quantum homomorphism to $Y$.

There is another graph parameter, $\chi_\text{vec}$, known as the \emph{vector chromatic number} \cite{chivec}, which is closely related to $\bartheta$. The only difference is that in the definition of $\chi_\text{vec}$, the vectors assigned to adjacent vertices must have inner product at most some $\alpha < 0$, as opposed to requiring equality. The analog of the theorem above for $\chi_\text{vec}$ holds, and the same proof works with the exception of changing the second '=' in Equation (\ref{eq:theta}) to '$\le$', and noting that $\tr(E_{xy}E_{x'y'}) \ge 0$ since both matrices are positive semidefinite.

\section{Quantum Parameters}\label{parameters}

There are several graph parameters that can be defined in terms of homomorphisms. Having defined quantum homomorphisms, it is natural to define quantum analogs of such graph parameters by simply replacing ``homomorphism'' with ``quantum homomorphism" in the definition. Here we consider the following:
\[
\begin{array}{ll}
\text{quantum chromatic number: } & \chi_q(X) := \min\{n \in \mathbb{N} : X \qarrow K_n\}; \\
\text{quantum clique number: } & \omega_q(X) := \max\{n \in \mathbb{N} : K_n \qarrow X\}; \\
\text{quantum independence number: } & \alpha_q(X) := \omega_q(\overline{X}); \\
\text{quantum odd girth: } & \og_q(X) := \min\{n \in \mathbb{N}, \ n \text{ odd} : C_n \qarrow X\}.
\end{array}
\]
The above definitions of quantum clique and independence numbers are different from those given in \cite{Beigi} and \cite{Winter}. The quantum clique number of \cite{Beigi} and the various independence numbers of \cite{Winter} are defined in terms of the amount of quantum or classical information one can send over a quantum channel, and are therefore more analogous to capacities than our notion of quantum independence number.

It is known that there exist graphs $X$ such that $\chi_q(X) < \chi(X)$ \cite{Avis,qchrom,qchrom-aqis,SS12,MSS}; we will see that there can also be (large) separations between $\alpha(X)$ and $\alpha_q(X)$ (and thus between $\omega(X)$ and $\omega_q(X)$). In contrast to these separations, we will show in Section~\ref{sec:oddgirth} that quantum odd girth is always equal to odd girth.

\subsection{Quantum Chromatic Number}

The quantum chromatic number has been relatively well-studied \cite{Avis,qchrom,qchrom-aqis,SS12,MSS}. In particular, a family of graphs exhibiting an exponential separation between $\chi$ and $\chi_q$ is known \cite{BCT99,BCW98,Avis}. In \cite{qchrom}, the authors show that when restricted to using rank one projectors, larger than exponential separations cannot be achieved. However, in \cite{qchrom-aqis} it is shown that rank one projectors are not always sufficient to attain $\chi_q(X)$. Despite these advances, some fundamental questions regarding quantum chromatic number remain unsolved. For example, it is not known whether $\chi_q(X)$ is computable or whether there exists a sequence of graphs $X_n$ for which $\lim_{n \rightarrow \infty} \chi(X_n) = \infty$ but $\lim_{n \rightarrow \infty} \chi_q(X_n) < \infty$. 

Our main result concerning quantum chromatic number is the following lower bound which is a direct consequence of Theorem~\ref{thetapreserved}:

\begin{cor}\label{chiqlowerbound}
For any graph $X$,
\[\bartheta(X) \le \chi_q(X).\]
\end{cor}

\proof
Let $c = \chi_q(X)$. Then $X \qarrow K_c$ and by Theorem~\ref{thetapreserved}, $\bartheta(X) \le \bartheta(K_c) = c$.\qed

This lower bound on $\chi_q(X)$ is new\footnote{Through private communication we have learned that $\vartheta(\overline{X}) \le \chi_q(X)$ has been independently proven (but not published) by M.~Laurent, G.~Scarpa, and A.~Varvitsiotis.} and, as $\omega(X) \le \bartheta(X)$, it improves upon the best previously known lower bound, $\omega(X)$. Furthermore, since $\omega(X)$ is NP-hard to compute, the above is also an improvement in terms of efficiency. This is particularly useful since it is not known if $\chi_q(X)$ is even computable.

We now use this lower bound to compute the exact value of the quantum chromatic number for a well-known class of graphs. For $n \in \mathbb{N}$, let $\Omega_n$ be the graph whose vertices are the $\pm 1$ vectors of length $n$ with orthogonal vectors being adjacent. We only consider the case when $4 | n$, since otherwise $\Omega_n$ is either empty or bipartite. In such a case, a result of Frankl and R\"{o}dl \cite{frankrod} implies that $\chi(\Omega_n)$ is exponential in $n$. On the other hand, it is known that $\chi_q(\Omega_n) \le n$ for all $n \in \mathbb{N}$ \cite{Avis}. However, it it remained unknown whether this inequality is tight for $4 | n$. To show this, we first compute $\bartheta(\Omega_n)$ for $4 | n$.

\begin{lemma}\label{thetahadamard}
If $n$ is divisible by 4, then $\bartheta(\Omega_n) = n$.
\end{lemma}
\proof\hspace{-.07in}\footnote{This proof is a minor modification of a proof from unpublished notes of Simone Severini.}
Let $n = 4m$. In $\cite{lovasz}$, it was shown that if $X$ is regular (all vertices have equal degree) and edge transitive, then
\[\vartheta(X) = |V(X)|\frac{\lambda_{\min}}{\lambda_{\min} - \lambda_{\max}}\]
where $\lambda_{\max}$ and $\lambda_{\min}$ are the largest and smallest eigenvalues of the adjacency matrix of $X$, respectively. Since $\Omega_n$ is regular and edge transitive, we can use this equality to compute $\vartheta(\Omega_n)$. The maximum eigenvalue of a regular graph is just the degree of the graph, and so for $\Omega_n$ this is equal to $\binom{4m}{2m}$. The minimum eigenvalue of $\Omega_n$ for $4|n$ was computed in \cite{interestinggraphs} and is equal to $-\binom{4m}{2m}/(4m-1)$. Therefore,
\[\vartheta(\Omega_n) = 2^n\frac{-\frac{\binom{4m}{2m}}{4m-1}}{-\frac{\binom{4m}{2m}}{4m-1}-\binom{4m}{2m}} = 2^n\frac{1}{1+(4m-1)} = \frac{2^n}{n}.\]
Another result from \cite{lovasz} states that for a vertex transitive graph $X$, \[\vartheta(X)\bartheta(X) = |V(X)|.\]
Since $\Omega_n$ is vertex transitive as well, we obtain that
\[\bartheta(\Omega_n) = n\]
as desired.\qed

Combining the above with Corollary~\ref{chiqlowerbound} yields the following:
\begin{cor}
If $n$ is divisible by 4, then $\chi_q(\Omega_n) = n$.\qed
\end{cor}

Curiously, $\{\Omega_{4n}\}_{n \in \mathbb{N}}$ is an infinite family of graphs for which the quantum chromatic number is known exactly, while the chromatic number remains unknown (for $n > 2$).

\subsection{Quantum Independence and Clique Numbers}\label{alpha_q}

In this section we show that any graphs $X$ and $Y$ such that $X \qarrow Y$ but $X \not\rightarrow Y$ can be used to construct a graph $Z$ for which $\alpha_q(Z) > \alpha(Z)$, and thus $\omega_q(\overline{Z}) > \omega(\overline{Z})$. Such constructions show that quantum advantage can also occur for homomorphism games with non-complete target graphs. We begin with a definition from \cite{Nesetril}:

\begin{definition}
For graphs $X$ and $Y$, define their \emph{homomorphic product}, denoted $X \ltimes Y$, to be the graph with vertex set $V(X) \times V(Y)$ with distinct vertices $(x,y)$ and $(x',y')$ being adjacent if either $x = x'$, or $x \sim x'$ and $y \not\sim y'$.
\end{definition}
It is useful to consider this product because the edges in $X \ltimes Y$ correspond exactly to the required orthogonalities for projectors giving a quantum homomorphism from $X$ to $Y$. We will often use the following two properties:
\begin{align}
\label{prop:omega} &K_{|V(Y)|} \rightarrow X \ltimes Y \\
\label{prop:chi} &\overline{X \ltimes Y} \rightarrow K_{|V(X)|}.
\end{align}
Both of these properties follow from the fact that the sets $V_x = \{(x,y): y \in V(Y)\}$ for $x \in V(X)$ induce cliques of size $|V(Y)|$ and partition $V(X \ltimes Y)$.

The following lemma is the original motivation for defining the homomorphic product, and it is stated as Exercise 7 in Chapter 2 of \cite{Nesetril}.

\begin{lemma}\label{lem:hom2indy}
For graphs $X$ and $Y$, we have that $X \rightarrow Y$ if and only if $\alpha(X \ltimes Y) = |V(X)|$.
\end{lemma}
\proof
Let $m = |V(X)|$ and $n = |V(Y)|$ and note that
\[\alpha(X \ltimes Y) = \omega(\overline{X \ltimes Y}) \le \chi(\overline{X \ltimes Y}) \le m\]
by Property~(\ref{prop:chi}). So $\alpha(X \ltimes Y) = m$ if and only if $X \ltimes Y$ has an independent set of size $m$.

Suppose that $\varphi : X \rightarrow Y$ is a homomorphism. Then the set $\{(x,\varphi(x)): x \in V(X)\}$ is an independent set of size $m$ in $X \ltimes Y$.

Conversely, let $S$ be an independent set of size $m$ in $X \ltimes Y$. Since the set $\{(x,y) : y \in V(Y)\}$ induces a clique for all $x \in V(X)$, and there are $m$ such sets partitioning $V(X\ltimes Y)$, there must be exactly one vertex of $S$ whose first coordinate is $x$ for every $x \in V(X)$. Define $\phi: V(X) \rightarrow V(Y)$ as follows:
\[\phi(x) = y \text{ for the unique } y \text{ such that } (x,y) \in S.\]
It is straightforward to check that $\phi$ is a homomorphism.\qed

Somewhat surprisingly, a quantum analog of this lemma also holds.
\begin{lemma}\label{lem:qhom2qindy}
For graphs $X$ and $Y$, we have that $X \qarrow Y$ if and only if $\alpha_q(X \ltimes Y) = |V(X)|$.
\end{lemma}
\proof
Let $m = |V(X)|$ and $n = |V(Y)|$. Note that
\[\alpha_q(X \ltimes Y) = \omega_q(\overline{X \ltimes Y}) \le \chi_q(\overline{X \ltimes Y}) \le \chi(\overline{X \ltimes Y}) \le m\]
by Property~(\ref{prop:chi}). So $\alpha_q(X \ltimes Y) = m$ if and only if $K_m \qarrow \overline{X \ltimes Y}$.

Suppose that $X \qarrow Y$ and consider Alice and Bob playing the $(K_m,\overline{X \ltimes Y})$-homomorphism game. We will describe a winning strategy for this game, thus proving one direction of the lemma. Since $m = |V(X)|$, we can assume that $K_m$ and $X$ have the same vertex set. When Alice and Bob are given inputs $\xa,\xb \in V(K_m) = V(X)$, they act as if they are playing the $(X,Y)$-homomorphism game with $\xa$ and $\xb$ as their inputs in order to obtain $\ya, \yb \in V(Y)$. They then output $(\xa,\ya)$ and $(\xb,\yb)$ respectively. If $\xa = \xb$, then $\ya = \yb$ and thus $(\xa,\ya) = (\xb,\yb)$. If $\xa \ne \xb$, then either $\xa \sim \xb$ in $X$ or $\xa \not\simeq \xb$ in $X$ (note that $\not\simeq$ means ``neither equal nor adjacent to" and is equivalent to ``adjacent in the complement"). In the former case, we will have that $\ya \sim \yb$ and thus $(\xa,\ya) \not\simeq (\xb,\yb)$ in $X \ltimes Y$ and therefore $(\xa,\ya) \sim (\xb,\yb)$ in $\overline{X \ltimes Y}$. In the latter case, we have that $(\xa,\ya) \not\simeq (\xb,\yb)$ in $X \ltimes Y$ regardless of $\ya$ and $\yb$, and therefore $(\xa,\ya) \sim (\xb,\yb)$ in $\overline{X \ltimes Y}$. So Alice and Bob can win the $(K_m,\overline{X \ltimes Y})$-homomorphism game and therefore $\alpha_q(X \ltimes Y) = |V(X)|$.

Conversely, suppose that $\alpha_q(X \ltimes Y) = m$, and that $P_{i(x,y)} \in \mathbb{M}_\mathbb{C}(d,d)$ for $i \in [m], \ x \in V(X), \ y \in V(Y)$ are the projectors corresponding to a winning strategy for the $(K_m,\overline{X \ltimes Y})$-homomorphism game. Then for all $i \in [n]$
\[\sum_{x \in V(X), y \in V(Y)} P_{i(x,y)} = I\]
and
\[P_{i(x,y)}P_{j(x',y')} = 0\]
whenever $\left(i = j \ \& \ (x,y) \ne (x',y')\right)$ OR $\left(i \ne j \ \& \ (x,y) \not\sim (x',y') \text{ in } \overline{X \ltimes Y}\right)$. Considering how adjacency is defined in $X \ltimes Y$, we can rewrite the orthogonality conditions above as $P_{i(x,y)}P_{j(x',y')} = 0$ if one of the following hold:
\begin{enumerate}[a.]
\item
$i = j$ and $(x,y) \ne (x',y')$, or
\item
$i \ne j$ and $x = x'$, or 
\item
$i \ne j$ and $x \sim x'$ and $y \not\sim y'$.
\end{enumerate}
For each $x \in V(X), y \in V(Y)$ define
\[Q_{xy} = \sum_{i \in [m]} P_{i(x,y)}.\]
We claim that the matrices $Q_{xy}$ are projectors which satisfy the conditions of Corollary \ref{cor:reformulation} for the $(X,Y)$-homomorphism game. They are indeed projectors since by (b) they are the sum of mutually orthogonal projectors.

Now we must check that the projectors $Q_{xy}$ satisfy the orthogonality conditions (\ref{cond:orthog}):
\begin{align*}
Q_{xy}Q_{x'y'} &= \left(\sum_{i \in [m]} P_{i(x,y)}\right)\left(\sum_{j \in [m]} P_{j(x',y')}\right) = \sum_{i,j \in [m]} P_{i(x,y)}P_{j(x',y')} \\
&= \left\{
     \begin{array}{ll}
       0 & \text{if } x = x' \ \& \ y \ne y' \text{ by (a) and (b)} \\
       0 & \text{if } x \sim x' \ \& \ y \not\sim y' \text{ by (a) and (c)} \\
       \text{something} & \text{o.w.}
     \end{array}
   \right.
\end{align*}
These are exactly the needed orthogonalities and so it is only left to show that for all $x \in V(X)$
\begin{equation}\label{eq:Q}
\sum_{y \in V(Y)} Q_{xy} = I.
\end{equation}
Since for a fixed $x \in V(X)$, the projectors $Q_{xy}$ are mutually orthogonal, we have that $\sum_{y \in V(Y)} \rk(Q_{xy}) \le d$ with equality if and only if (\ref{eq:Q}) holds. By (a) we have that
\[\sum_{x \in V(X), y \in V(Y)} \rk\left(P_{i(x,y)}\right) = \rk\left(\sum_{x \in V(X), y \in V(Y)} P_{i(x,y)}\right) = \rk(I) = d\]
for all $i \in [m]$. Therefore,
\begin{align*}
md &= \sum_{i \in [m]}\left(\sum_{x \in V(X), y \in V(Y)} \rk\left(P_{i(x,y)}\right)\right) = \sum_{x \in V(X), y \in V(Y)} \rk\left(\sum_{i \in [m]}P_{i(x,y)}\right) \\
&= \sum_{x \in V(X)} \left(\sum_{y \in V(Y)} \rk\left(Q_{xy}\right)\right) \le |V(X)|d = md
\end{align*}
where the second equality follows from (b). Of course this implies that $\sum_{y \in V(Y)} \rk(Q_{xy}) = d$ for all $x \in V(X)$ and thus (\ref{eq:Q}) holds and we are done.\qed

Interestingly, one direction of the above lemma can be proved without appealing to the actual strategy being used, while the other direction cannot. Indeed, if $X$ is the path on three vertices, and $Y = K_1$, then there is no non-signalling correlation which wins the $(X,Y)$-homomorphism game (if Alice and Bob receive adjacent vertices they have no way of winning). However, $\overline{X \ltimes Y}$ is a single edge plus an isolated vertex, and so any graph (including arbitrarily large cliques) can be mapped to $\overline{X \ltimes Y}$ using a non-signalling correlation as described in Section~\ref{nonsignal}. Therefore the non-signalling version of the ``if" direction of this lemma does not hold, and thus any proof of it has to appeal to the particular type of strategy being used.

We now show how to construct graphs exhibiting separations between independence and quantum independence numbers. This construction is inspired by, and a generalization of, Theorem 16 of \cite{MSS}.

\begin{cor}
If $X \not\rightarrow Y$ and $X \qarrow Y$, then $\alpha(X \ltimes Y) < \alpha_q(X \ltimes Y)$.
\proof
If $X \not\rightarrow Y$, then $\alpha(X \ltimes Y) < |V(X)|$. If $X \qarrow Y$, then $\alpha_q(X \ltimes Y) = |V(X)|$.\qed
\end{cor}

By the definition of quantum independence number, the above corollary states that if $X \not\rightarrow Y$ and $X \qarrow Y$, then $K_n \not\rightarrow \overline{X\ltimes Y}$ and $K_n \qarrow \overline{X\ltimes Y}$, where $n = |V(Y)|$. Therefore the $(K_n,\overline{X \ltimes Y})$-homomorphism game exhibits quantum advantage, and by Lemma~\ref{complete} the graph $\overline{X\ltimes Y}$ is not (homomorphically equivalent to) a complete graph.

Unfortunately, the above result does not guarantee a large separation. However, we will see how to remedy this in some cases.

For graphs $X$ and $Y$, their Cartesian product, denoted $X\square Y$, is the graph with vertex set $V(X) \times V(Y)$ with two vertices being adjacent if they are equal in one coordinate and adjacent in the other. Note that $X \ltimes K_n = X \square K_n$ for all graphs $X$. A classic theorem of Vizing's states:
\[\alpha(X \square Y) \le \min\{\alpha(X)|V(Y)|, \alpha(Y)|V(X)|\}.\]
Suppose that $X$ is a graph such that $\chi_q(X) = c \ll |V(X)|/\alpha(X)$ (note that this necessarily implies that $\chi_q(X) \ll \chi(X)$). We have that $X \qarrow K_c$ and therefore $\alpha_q(X \square K_c) = |V(X)|$. However, applying the result of Vizing, we obtain $\alpha(X \square K_c) \le c\alpha(X) \ll |V(X)|$.

Applying this approach to the graphs $\Omega_n$, we get the following:

\begin{cor}\label{qindyexample}
There exists an $\epsilon > 0$, such that for $n \in \mathbb{N}$ divisible by 4
\[\frac{\alpha_q(\Omega_n \square K_n)}{\alpha(\Omega_n \square K_n)} \ge \frac{1}{n}\left(\frac{2}{2-\epsilon}\right)^n.\]
\end{cor}
\proof
We have mentioned that $\Omega_n \qarrow K_n$, so by Lemma \ref{lem:qhom2qindy}, $\alpha_q(\Omega_n \square K_n) = |V(\Omega_n)| = 2^n$. A result by Frankl and R\"{o}dl \cite{frankrod} shows that there exists an $\epsilon > 0$ such that for $n \in \mathbb{N}$ divisible by 4, $\alpha(\Omega_n) \le (2 - \epsilon)^n$. Applying the Vizing result gives that $\alpha(\Omega_n \square K_n) \le n(2-\epsilon)^n$.
\qed

The idea and proof for the above corollary is identical to Theorem 17 in \cite{MSS}, though their result is about one-shot zero-error capacities rather than independence numbers.\\

As with quantum chromatic number, Theorem~\ref{thetapreserved} lets us bound the quantum clique number of a graph.

\begin{cor}\label{omegaqupperbound}
For any graph $X$,
\[\omega_q(X) \le \bartheta(X) \ \ \& \ \ \alpha_q(X) \le \vartheta(X).\]
\end{cor}
\proof
Let $n = \omega_q(X)$. Then $K_n \qarrow X$, and by Theorem~\ref{thetapreserved} we have $n = \bartheta(K_n) \le \bartheta(X)$.\qed

Applying the above to Lemma~\ref{lem:qhom2qindy}, we further obtain:

\begin{cor}\label{thetaproduct}
If $X \qarrow Y$, then
\[\vartheta(X \ltimes Y) = |V(X)|.\]
\end{cor}
\proof
First, we have that
\[\vartheta(X \ltimes Y) = \bartheta(\overline{X \ltimes Y}) \le \chi(\overline{X \ltimes Y}) \le |V(X)|,\]
where the last inequality follows by Property~(\ref{prop:chi}) of the homomorphic product.
Suppose that $X \qarrow Y$, then by Corollary~\ref{omegaqupperbound} and Lemma~\ref{lem:qhom2qindy} we have
\[\vartheta(X \ltimes Y) \ge \alpha_q(X \ltimes Y) = |V(X)|.\]\qed

Though this corollary is weaker than Lemma~\ref{lem:qhom2qindy}, it may be of practical use since $\vartheta$ can be approximated efficiently.

Another nice consequence of Theorem~\ref{thetapreserved}, and more specifically Corollary~\ref{omegaqupperbound}, is a quantum version of the well-known clique-coclique bound for vertex transitive graphs:

\begin{cor}\label{cliquecoclique}
If $X$ is a vertex transitive graph, then
\[\alpha_q(X) \omega_q(X) \le |V(X)|.\]
\end{cor}
\proof
In \cite{lovasz}, it is shown that $\vartheta(X)\bartheta(X) = |V(X)|$ for any vertex transitive graph $X$. Applying Corollary~\ref{omegaqupperbound} we see that
\[\alpha_q(X)\omega_q(X) \le \vartheta(X)\bartheta(X) = |V(X)|.\]\qed

The clique-colique bound is a special case of the no-homomorphism lemma \cite{nohomo}, which states that if $X \rightarrow Y$, and $Y$ is vertex transitive, then
\[\frac{|V(X)|}{\alpha(X)} \le \frac{|V(Y)|}{\alpha(Y)}.\]
So it is natural to ask if a quantum version of the no-homomorphism lemma holds true. In particular this would imply that for all graphs $X$,
\[\frac{|V(X)|}{\alpha_q(X)} \le \chi_q(X).\]
It turns out that this does not hold. Recalling the graph $\Omega_n$, for $4 | n$ we have the following:
\[\chi_q(\Omega_n) = n \ \ \text{and} \ \ \alpha_q(\Omega_n) \le \vartheta(\Omega_n) = \frac{2^n}{n} \ \ \text{and} \ \ |V(\Omega_n)| = 2^n.\]
Since, $\alpha_q(\Omega_n)$ is integer, we have that $\alpha_q(\Omega_n) \le \lfloor \frac{2^n}{n} \rfloor$, and thus $\alpha_q(\Omega_n) < \frac{2^n}{n}$ whenever $n$ is not a power of 2. Therefore, we have that
\[\frac{|V(\Omega_n)|}{\alpha_q(\Omega_n)} > \frac{2^n}{\frac{2^n}{n}} = n = \chi_q(\Omega_n)\]
when $n$ is a multiple of 4 but not a power of 2.

\subsection{Walks and Quantum Odd Girth}
\label{sec:oddgirth}

In this section we show that, unlike for quantum chromatic and independence numbers, there can be no separation between odd girth and quantum odd girth. The key ingredient in this proof is Lemma~\ref{lem:walks} below which concerns walks in graphs $X$ and $Y$ such that $X \qarrow Y$. A \emph{walk} of length $\ell$ from $x$ to $x'$ in a graph $X$ is a sequence of vertices $x = x_0, x_1, \ldots, x_\ell = x'$ such that $x_{i-1} \sim x_{i}$ for $i \in [\ell]$. If $x = x'$, then we say the walk is \emph{closed}. Note that both vertices and edges are allowed to be repeated in a walk. We will use $x \sim_\ell x'$ to denote that there exists a walk of length $\ell$ from $x$ to $x'$.

Suppose that $\varphi$ is a homomorphism from $X$ to $Y$. For any walk $x_0, x_1, \ldots, x_\ell$ in $X$, we have that $\varphi(x_{i-1}) \sim \varphi(x_i)$ for all $i \in [\ell]$, since homomorphisms preserve adjacency. Therefore, $\varphi(x) \sim_\ell \varphi(x')$ whenever $x \sim_\ell x'$. Lemma~\ref{lem:walks} below is a quantum analog of this well-known classical result. In terms of the homomorphism game, Lemma~\ref{lem:walks} says that, for any winning quantum strategy, if there is a walk of length $\ell$ between Alice and Bob's input vertices, then they always respond with vertices which have a walk of length $\ell$ between them.


\begin{lemma}\label{lem:walks}
Let $X$ and $Y$ be graphs and suppose that $\{E_{xy} : x \in V(X), y \in V(Y)\}$ are projectors which give a quantum homomorphism from $X$ to $Y$. If $x \sim_\ell x'$ and $y \not\sim_\ell y'$, then $E_{xy}E_{x'y'} = 0$.
\end{lemma}
\proof
We use induction on the length of the walk $\ell$. For $\ell \in \{0,1\}$ the lemma holds by Condition~(\ref{cond:orthog}). Now suppose that it holds for some $\ell \ge 1$. Let $x$ and $x'$ be vertices of $X$ such that $x = x_0, x_1, \ldots, x_\ell, x_{\ell+1} = x'$ is a walk of length $\ell+1$ in $X$. Let $y,y' \in V(Y)$ be vertices such that $y \not\sim_{\ell+1} y'$. Since there exists a walk of length $\ell$ from $x$ to $x_\ell$, by our induction hypothesis we have that
\[E_{xy}E_{x_\ell y''} = 0\]
for all $y''$ such that $y \not\sim_\ell y''$. Therefore,
\[E_{xy} \sum_{y'' : y \sim_\ell y''} E_{x_\ell y''} = E_{xy}.\]
Thus, it suffices to show that
\[E_{x'y'} \sum_{y'' : y \sim_\ell y''} E_{x_\ell y''} = 0.\]
Note that if $y \sim_\ell y''$ and $y'' \sim y'$, then $y \sim_{\ell+1} y'$ which is a contradiction. Therefore, no such $y''$ exists. From this we have that
\[E_{x'y'} \sum_{y'' : y \sim_\ell y''} E_{x_\ell y''} = E_{x'y'} \sum_{y'' : y \sim_\ell y'' \ \& \ y' \not\sim y''} E_{x_\ell y''} = 0,\]
since $x_\ell \sim x'$.\qed

Using Lemma~\ref{lem:walks}, the equality of odd girth and quantum odd girth is almost immediate:

\begin{theorem}\label{thm:oddgirth}
For any graph $X$ and integer $g \in \mathbb{N}$,
\[C_g \qarrow X \text{ if and only if } C_g \to X.\]
In particular, this implies that
\[\og_q(X) = \og(X)\]
for any graph $X$.
\end{theorem}
\proof
For even $g$, the graph $C_g$ is bipartite and therefore homomorphically equivalent to $K_2$. Therefore the statement $C_g \to X$ is equivalent to $X$ having an edge. It is easy to see that $K_2 \qarrow X$ if and only if $X$ has an edge, and thus the theorem holds for even $g$.

Suppose that $g$ is odd and $C_g \qarrow X$. Let $E_{ux}$ for $u \in V(C_g)$ and $x \in V(X)$ be the projectors that give this quantum homomorphism. For any $v \in V(C_g)$, there exists a walk of length $g$ in $C_g$ beginning and ending at $v$, thus $v \sim_g v$. By Condition~(\ref{cond:ident}), there must exist an $x' \in V(X)$ such that $E_{vx'} \ne 0$. Therefore, $E_{vx'}E_{vx'} \ne 0$ and by Lemma~\ref{lem:walks} we have that $x' \sim_g x'$. In other words, $X$ has a closed walk of \emph{odd} length $g$. This is only possible if $X$ contains an odd cycle of length $g$ or less, and thus $C_g \to X$. Finally, since $C_g \to X$ implies $C_g \qarrow X$, we have proved the first claim. The second claim follows directly from the definitions of odd girth and quantum odd girth.\qed

We also have the following corollary for small quantum clique number:

\begin{cor}
If $\omega_q(X) = 3$, then $\omega(X) = 3$.
\end{cor}
\proof
This follows from Theorem~\ref{thm:oddgirth} and the fact that $C_3 = K_3$.\qed

\subsection{Quantum Homomorphisms to Kneser Graphs}

Here we prove a lemma which relates the existence of quantum homomorphism from $X$ to $Y$ to the existence of a quantum homomorphism from $X \ltimes Y$ to a particular Kneser graph. The Kneser graph, $K_{n:r}$, is the graph whose vertices are the $r$-subsets of the set $[n]$ with disjoint sets being adjacent.

\begin{lemma}\label{kneser2hom}
Let $X$ and $Y$ be graphs. If $n,r \in \mathbb{N}$ are such that $\frac{n}{r} = |V(Y)|$, then 
\[X \ltimes Y \rightarrow K_{n:r} \Rightarrow X \rightarrow Y\]
and 
\[X \ltimes Y \qarrow K_{n:r} \Rightarrow X \qarrow Y\]
\end{lemma}
\proof
Suppose that $X \ltimes Y \rightarrow K_{n:r}$. The $r$-subsets of $[n]$ containing the element 1 give an independent set of $K_{n:r}$ of size $\binom{n-1}{r-1}$, and thus $\alpha(K_{n:r}) \ge \binom{n-1}{r-1}$ for all values of $n$ and $r$. 
Since $K_{n:r}$ is vertex transitive, we can apply the no-homomorphism lemma \cite{nohomo} to obtain that
\[\frac{|V(X\ltimes Y)|}{\alpha(X\ltimes Y)} \le \frac{|V(K_{n:r})|}{\alpha(K_{n:r})} \le \frac{n}{r} = |V(Y)|.\]
Therefore, $\alpha(X \ltimes Y) \ge |V(X)|$ and by Lemma~\ref{lem:hom2indy} $X \rightarrow Y$.

The proof of the quantum statement uses results and discussion from Section~\ref{rank}. Suppose that $X \ltimes Y \qarrow K_{n:r}$. Since $K_{n:r}$ has a projective representation of value $\frac{n}{r} = |V(Y)|$, by Theorem~\ref{thm:xipreserved} so does $X \ltimes Y$, and therefore by Corollary~\ref{hom2rep}, $X \qarrow Y$.\qed

\begin{lemma}\label{hom2kneser}
If $X$ and $Y$ are graphs and $G$ is a subgroup of $\text{Aut}(Y)$ that acts transitively on $V(Y)$, then
\[X \rightarrow Y \Leftrightarrow X \ltimes Y \rightarrow K_{|G|:\frac{|G|}{|V(Y)|}}\]
and 
\[X \qarrow Y \Leftrightarrow X \ltimes Y \qarrow K_{|G|:\frac{|G|}{|V(Y)|}}\]
\end{lemma}
\proof
Let $G$ be as stated and let $n = |G|$ and $r = |G|/|V(Y)|$. Note that we can view the vertices of $K_{n:r}$ as being $r$-subsets of $G$. We will prove both forward implications simultaneously. Suppose that Alice and Bob can win the $(X,Y)$-homomorphism game. For all $y,y' \in V(Y)$ let
\[G(y,y') = \{\phi \in G : \phi(y) = y'\}.\]
Note that $G(y,y')$ is a coset of the stabilizer of $y$. Since $G$ acts transitively on $V(Y)$, the orbit of any vertex is $V(Y)$, and therefore
\[|G(y,y')| = |G|/|V(Y)| = r\]
for all $y,y' \in V(Y)$. Consider the following strategy for the $(X \ltimes Y,K_{n:r})$-homomorphism game. Upon receiving $(x_A,y_A), (x_B,y_B) \in V(X\ltimes Y)$, Alice and Bob act as if they are playing the $(X,Y)$-homomorphism game with $x_A, x_B \in V(X)$ to obtain $y'_A, y'_B \in V(Y)$. They then respond with $G(y_A,y'_A)$ and $G(y_B,y'_B)$ respectively. To show that this is a valid strategy for the $(X \ltimes Y,K_{n:r})$-homomorphism game, we have two cases:

\begin{itemize}
\item If $(x_A,y_A) = (x_B,y_B)$, then $x_A = x_B$ and therefore $y'_A = y'_B$. Furthermore, $y_A = y_B$ and so $G(y_A,y'_A) = G(y_B,y'_B)$.

\item If $(x_A,y_A) \sim (x_B,y_B)$, then we have two subcases:
\begin{itemize}
\item If $x_A = x_B$ and $y_A \ne y_B$, then $y'_A = y'_B$. Since no automorphism of $Y$ can map both $y_A$ and $y_B$ to the same vertex, $G(y_A,y'_A)$ and $G(y_B,y'_B)$ are disjoint.
\item If $x_A \sim x_B$ and $y_A \not\sim y_B$, then $y'_A \sim y'_B$. Since no automorphism of $Y$ can map two (possibly equal) nonadjacent vertices to two adjacent vertices, $G(y_A,y'_A)$ and $G(y_B,y'_B)$ are disjoint.
\end{itemize}
\end{itemize}

The other direction of each statement is proven by Lemma~\ref{kneser2hom} and the fact that
\[\frac{|G|}{|G|/|V(Y)|} = |V(Y)|.\]\qed

Applying Theorem~\ref{thetapreserved} to the above, we obtain a corollary similar to Corollary~\ref{thetaproduct}:

\begin{cor}\label{thetaproducttrans}
If $Y$ is vertex transitive and $X \qarrow Y$, then
\[\bartheta(X \ltimes Y) = |V(Y)|.\]
\end{cor}
\proof
Recalling Property~(\ref{prop:omega}), we see that
\[\bartheta(X \ltimes Y) \ge \omega(X \ltimes Y) \ge |V(Y)|.\]
By Lemma~\ref{hom2kneser}, $X \ltimes Y \qarrow K_{n:r}$ for $n = |\text{Aut}(Y)|$, and $r = |\text{Aut}(Y)|/|V(Y)|$. Applying Theorem~\ref{thetapreserved} we have that
\[\bartheta(X \ltimes Y) \le \bartheta(K_{n:r}) = \frac{n}{r} = |V(Y)|\]
where the value of $\bartheta(K_{n:r})$ comes from \cite{lovasz}.\qed

As with Corollary~\ref{thetaproduct}, the above result is strictly weaker than Lemma~\ref{hom2kneser}, but since $\bartheta$ is efficient to approximate, it may be of more practical use. We note that in the case of $Y$ being vertex transitive, it has been shown in~\cite{conic} that $ \bartheta(X \ltimes Y) = |V(Y)|$ is equivalent to $\vartheta(X \ltimes Y) = |V(X)|$. Therefore, Corollary~\ref{thetaproducttrans} is in fact implied by Corollary~\ref{thetaproduct}.


\section{Zero-Error Capacity}\label{capacity}

In this section we investigate the relationship of quantum
homomorphisms to zero-error channel capacity. We start by explaining
the setting of zero-error communication.

Suppose Alice wants to use a noisy classical channel $\mathcal{N}$ to
communicate messages to Bob with zero probability of decoding error.
The channel $\mathcal{N}$ is completely specified by the set of inputs
$Z$, the set of outputs $W$ and a conditional probability distribution
$\mathcal{N}(w|z)$, the probability to output $w$ upon inputting
$z$. Inputs $z$ and $z'$ are said to be \emph{confusable} if there
exists $w\in W$ such that $\mathcal{N}(w|z) > 0$ and
$\mathcal{N}(w|z') > 0$. To represent a channel $\mathcal{N}$, we use
its \emph{confusability graph} $X_{\mathcal{N}}$. This is a graph with
vertex set $Z$ in which distinct vertices $z,z'$ are adjacent whenever
inputs $z$ and $z'$ are confusable. Note that $X_{\mathcal{N}}$ does
not have loops even though input $z$ is confusable with itself.

Let us identify the messages that Alice wants to transmit to Bob with the integers from $[m]$. To send a messages, the parties need to agree on encoding function $\mathcal{E}:[m]\to Z$, which Alice uses to map messages to inputs on the channel. They also need a decoding function $\mathcal{D}:W\to [m]$, which Bob uses to map outputs of the channel to messages. Since, we are interested in zero-error communication, we must have that for all $i\in [m]$ 
\begin{equation}
  \mathcal{D}(w)=i \quad 
  \text{for all $w$ with $\mathcal{N}(w|\mathcal{E}(i))>0$.}
\label{eq:ZECond}
\end{equation}
If the channel inputs $z=\mathcal{E}(i)$ and $z'= \mathcal{E}(i')$ are confusable for some $i\neq i'$, then it is impossible for Bob to choose $\mathcal{D}$ so that Condition~(\ref{eq:ZECond}) is satisfied. This is because, there exists an output $w\in W$ which can be produced upon both of the inputs $z$ and $z'$. So in case Bob receives $w$ he will not be able to tell if Alice intended to send message $i$ or $i'$. Therefore the image of encoding function $\mathcal{E}$ must consist entirely of mutually unconfusable inputs. In other words, the image of $\mathcal{E}$ is an independent set in the confusability graph $X_{\mathcal{N}}$. On the other hand, it is also easy to see that if the image of $\mathcal{E}$ forms an independent set of $X_{\mathcal{N}}$, then decoding is always possible.

The one-shot zero-error capacity, $c_0(\mathcal{N})$, is the largest
$m$ for which Alice and Bob can choose the encoding and decoding functions which allow them to transmit any of the $m$ messages with zero chance of error through one use of $\mathcal{N}$. From the discussion in the previous paragraph it is clear that $c_0(\mathcal{N})=\alpha(X_{\mathcal{N}})$.
It turns out that one-shot zero-error capacity can increase if Alice
and Bob are allowed to use shared entanglement \cite{entcap}. The capacity in this case is referred to as the entanglement-assisted one-shot zero-error capacity, and denoted $c_0^*(\mathcal{N})$. As in the unassisted case, it can be shown that
$c_0^*(\mathcal{N})$ only depends on the confusability graph
$X_{\mathcal{N}}$ \cite{entcap}. Thus, from now on we can consider one-shot capacities of graphs and write $c_0(X)$ or $c_0^*(X)$. For
each graph $X$, we choose a canonical channel $\mathcal{N}$ that has
$X$ as its confusability graph. This is the channel with input set
$V(X)$, output set $E(X)$, and 
\begin{equation}
  \mathcal{N}(e|x)=\begin{cases}
                     \frac{1}{deg(x)} & \text{if $e$ is incident to $x$}\\
                     0 & \text{if otherwise.}\\
				   \end{cases}
\end{equation}
For isolated vertices we can either add loops or allow for isolated
vertices to be outputs of our channel.

We now describe the general form of a quantum protocol that Alice
and Bob can use to transmit one of $m$ different messages over
$\mathcal{N}$ using a shared entangled state $\psi$. To send message
$i \in [m]$, Alice measures her part of $\psi$ using a POVM
$\mathcal{E}_i = \{E_{iz}\}_{z \in Z}$. Alice gets some outcome $z$ which leaves Bob with unnormalized residual state $\beta_{iz} := \tr_A((E_{iz} \otimes I)\psi\psi^*)$, where $\tr_A $ is the partial trace (see Section~\ref{sec:Prelim}).  We note that $\beta_{iz}=0$ if and only if the probability of Alice obtaining outcome $z$ is zero.
Alice then uses the measurement outcome $z$ as the input to the channel. Bob receives some output $w$
such that $\mathcal{N}(w|z)>0$. To recover the original message $i$, Bob performs a measurement on his residual state $\beta_{iz}$. We will explain that, as was shown in~\cite{entcap}, a successful recovery can be ensured if and only if
\begin{equation}
  \tr(\beta_{iz} \beta_{jz'}) = 0 \text{ for all distinct $i,j\in[m]$ and
all confusable $z,z'\in Z$.}
\label{eq:betas}
\end{equation}
First, note that $\tr(\beta_{iz} \beta_{jz'}) = 0$ if and only if $\beta_{iz}\beta_{jz'} = 0$ since $\beta_{iz}$ and $\beta_{jz'}$ are positive semidefinite. 
To show that Condition~(\ref{eq:betas}) is necessary, suppose for contradiction that for some $i,j \in [m]$ and confusable $z,z' \in Z$ we have $\tr(\beta_{iz} \beta_{jz'} )\ne 0$. In such case, $\beta_{iz}, \beta_{jz'} \ne 0$ and hence Alice obtains outcome $z$ upon measurement $\mathcal{E}_i$ with nonzero probability and similarly for $z'$ and $\mathcal{E}_j$. Since $z$ and $z'$ are confusable, there exists a $w \in W$ 
such that $\mathcal{N}(w|z) >0 $ and $\mathcal{N}(w|z') >0$. Therefore, it is possible for Bob to receive $w$ when Alice is trying to send $i$ and when she is trying to send $j$. Since Bob cannot use the channel output $w$ to recover Alice's messge, he must use his residual state. In the case where Alice was trying to send $i$, Bob has residual state $\beta_{iz}$, and in the case where she was trying to send $j$ he has $\beta_{jz'}$. 
In quantum information it is a well-known fact that only orthogonal quantum states can be perfectly distinguished. Since by assumption the states $\beta_{iz}$ and $\beta_{jz' }$ are not orthogonal, Bob cannot tell with certainty which state he has. Therefore, he cannot conclude with certainty whether Alice was trying to send message $i$ or $j$.

On the other hand, suppose that Condition~(\ref{eq:betas}) does hold. We will describe a strategy for Bob which lets him determine with certainty which message Alice was trying to send. For each $i \in [m]$ and $w \in W$, define $P_{wi}$ to be the projection onto the image (column space) of the following operator:
\[\alpha_{wi} := \sum_{z : \mathcal{N}(w|z)>0} \beta_{iz}.\]
For distinct messages $i,j$ and channel inputs $z,z'$ and output $w$ such that $\mathcal{N}(w|z), \mathcal{N}(w|z') > 0$, by Condition~(\ref{eq:betas}) we have that $\beta_{iz}$ and $\beta_{jz'}$ are orthogonal.
This implies that $\alpha_{wi}$ and $\alpha_{wj}$ are orthogonal for $i \ne j$, and therefore the projectors $P_{wi}$ and $P_{wj}$ are orthogonal for $i \ne j$. We can therefore define the projective measurement
\[\mathcal{F}_w := \{P_{wi} : i \in [m]\} \cup \left\{I - \sum_i P_{wi}\right\},\]
which Bob performs upon receiving output $w$. If Alice was trying to send message~$i$, then Bob's (normalized) residual state is $\beta'_{iz} := \frac{1}{\tr(\beta_{iz})}\beta_{iz}$ for some $z$ such that $\mathcal{N}(w|z) > 0$. Since the image of $\beta_{iz}$ is contained in the image of $\alpha_{wi}$, we have that $P_{wi}\beta_{iz} = \beta_{iz}$ and therefore $\tr(P_{wi}\beta'_{iz}) = \tr(\beta'_{iz}) = 1$. Thus Bob obtains outcome $i$ from his measurement with probability 1 and from this can correctly conclude that Alice was sending message $i$.

\subsection{Relation to Quantum Independence Number}

Recalling that $c_0(X) = \alpha(X)$ for any graph $X$, it is natural to ask if there is any relation between the entanglement-assisted one-shot capacity and the quantum independence number. The following theorem shows that these two parameters are indeed related.

\begin{theorem}\label{thm:qindyqcap}
For any graph $X$,
\[\alpha_q(X) \le c_0^*(X)\]
with equality if and only if $c_0^*(X)$ can be achieved using a strategy in which all of Alice's measurements are projective and the shared state is $\Phi = \frac{1}{\sqrt{d}}\sum_{i = 1}^d e_i \otimes e_i \in \mathbb{C}^d \otimes \mathbb{C}^d$ for some $d \in \mathbb{N}$.
\end{theorem}
\proof
We will make use of the following identity, which can be proved easily using the techniques from \cite{calculus}:
\[\tr(\tr_A((E \otimes I)\Phi \Phi^*)\tr_A((F \otimes I) \Phi \Phi^*)) = \frac{1}{d^2}\tr(EF).\]
Let $E_{ix} \in M_{\mathbb{C}}(d,d)$ be projectors that give a quantum homomorphism from $K_m$ to $\overline{X}$, where $m = \alpha_q(X)$. We will show that the projectors $E_{ix}$ along with the shared state $\Phi$ give a valid strategy for sending $m$ messages with $X$. Bob's residual states for this strategy are
\[\beta^i_x = \tr_A((E_{ix} \otimes I)\Phi \Phi^*).\]
It remains to show that the above operators satisfy Condition~(\ref{eq:betas}). If $i \ne j$ and ($x \sim x'$ or $x = x'$), then by the winning conditions of the $(K_m,\overline{X})$-homomorphism game, we have that $\tr(E_{ix}E_{jx'}) = 0$, and therefore
\[\tr(\beta^i_x \beta^j_{x'}) = 0.\]
That is, Condition~(\ref{eq:betas}) is satisfied. Therefore $\alpha_q(X) \le c_0^*(X)$.

Now suppose that Alice can use projective measurements $E_{ix}$ and the shared state, $\Phi$, to send $m$ messages with $X$. From the discussion above, we have that
\begin{align*}
0 = \tr(\tr_A((E_{ix} \otimes I)\Phi \Phi^*)\tr_A((E_{jx'} \otimes I) \Phi \Phi^*)) = \frac{1}{d^2}\tr(E_{ix}E_{jx'})
\end{align*}
for all $i \ne j$, and confusable $x,x'$. Note that $x$ is confusable with $x'$ if and only if $x \sim x'$ or $x = x'$. Since the matrices $E_{ix}$'s are projectors, we also have that $\tr(E_{ix}E_{ix'}) = 0$ for all $i \in [m]$ and $x \ne x'$. These are all the orthogonalities required to win the $(K_m,\overline{X})$-homomorphism game.
\qed

If in fact $c_0^*(X)$ can always be achieved using projective measurements on the maximally entangled state, then $\alpha_q(X) = c_0^*(X)$. In light of this we conjecture the following:
\begin{conjecture}\label{conj:qindyqcap}
For any graph $X$,
\[\alpha_q(X) = c_0^*(X).\]
\end{conjecture}

As a corollary to the above theorem, we obtain a method for constructing a graph $Z$ with $c_0(Z) < c_0^*(Z)$ given graphs $X$ and $Y$ such that $X \qarrow Y$ but $X \not\rightarrow Y$. This was already proven for $Y = K_n$ in \cite{MSS}.

\begin{cor}
If $X \qarrow Y$ but $X \not\rightarrow Y$, then
\[c_0(X \ltimes Y) < c_0^*(X \ltimes Y) = |V(X)|.\]
\end{cor}
\proof
First, note that
\[c_0^*(X\ltimes Y) \le \vartheta(X \ltimes Y) \le \chi(\overline{X\ltimes Y}) \le |V(X)|.\]
Now since $X \not\rightarrow Y$ and $X \qarrow Y$, we have that
\[c_0(X \ltimes Y) = \alpha(X \ltimes Y) < |V(X)| = \alpha_q(X\ltimes Y) \le c_0^*(X \ltimes Y) \le |V(X)|.\]\qed

\subsection{Relation to Quantum Homomorphisms}

Here we show that quantum homomorphisms respect the order of entanglement-assisted one-shot zero-error capacity.

\begin{lemma}
If $X \qarrow Y$, then $c_0^*(\overline{X}) \le c_0^*(\overline{Y})$.
\end{lemma}
\proof
Suppose that $\overline{X} \qarrow \overline{Y}$ and that $P_{xy} \in\mathbb{M}_\mathbb{C}(d,d)$, for $x \in V(X), y \in V(Y)$ are projectors which give such a quantum homomorphism. Furthermore, let $E_{ix}$, $i \in [m]$, $x \in V(X)$, be POVM elements and $\psi$ the shared state giving a strategy for sending $m$ messages with $X$. Then Bob's residual states for this strategy
\[\beta_x^i = \tr_A((E_{ix} \otimes I)\psi\psi^*)\]
satisfy the orthogonality condition~(\ref{eq:betas}).
Let $\Phi = \frac{1}{\sqrt{d}}\sum_{i = 1}^d e_i \otimes e_i$, and let
\[F_{iy} = \sum_{x \in V(X)} P_{xy} \otimes E_{ix}\]
for all $i \in [m]$, $y \in V(Y)$. We claim that the POVMs $\{F_{iy}\}_{y \in V(Y)}$ and the state $\Phi \otimes \psi$ give a valid strategy for sending $m$ messages with $Y$. First, note that
\[\sum_{y \in V(Y)} F_{iy} = \sum_{x \in V(X), y \in V(Y)} P_{xy} \otimes E_{ix} = \sum_{x \in V(X)} I \otimes E_{ix} = I\]
Second, each $F_{iy}$ is positive semidefinite since it is the sum of tensor products of positive semidefinite matrices, therefore the sets $\{F_{iy}\}_{y \in V(Y)}$ are indeed POVMs. If we use $A'$ and $A$ to denote the spaces in Alice's posession, then the the residual states are given by
\begin{align*}
\alpha_y^i &= \tr_{A'A}\left(\sum_{x \in V(X)}\left((P_{xy} \otimes I)\Phi\Phi^*\right) \otimes \left((E_{ix} \otimes I)\psi\psi^*\right)\right) \\
&= \frac{1}{d}\sum_{x \in V(X)} P_{xy} \otimes \tr_A((E_{ix} \otimes I)\psi\psi^*) = \frac{1}{d}\sum_{x \in V(X)} P_{xy} \otimes \beta_x^i.
\end{align*}
Suppose that $i \ne j$ and $y,y'$ are confusable, i.e.~$y$ is not adjacent to $y'$ in $\overline{Y}$. Then
\[
\alpha_y^i \alpha_{y'}^j = \frac{1}{d^2}\sum_{x,x' \in V(X)} P_{xy}P_{x'y'} \otimes \beta_x^i \beta_{x'}^j = \frac{1}{d^2}\sum_{x \not\simeq x' \hspace{.015in} \text{in} \hspace{.015in} X} P_{xy}P_{x'y'} \otimes \beta_x^i \beta_{x'}^j = 0
\]
where the last equality holds because of the orthogonality constraints on the projectors $P_{xy}$. This shows that Alice and Bob can send $m$ messages using $Y$.\qed

The entanglement-assisted Shannon capacity, $\Theta^*$, is the asymptotic version of $c_0^*$, and is defined as
\[\Theta^*(X) = \lim_{n \rightarrow \infty} \sqrt[n]{c_0^*(X^{\boxtimes n})}.\]
Here, $X \boxtimes Y$ is the graph with vertex set $V(X) \times V(Y)$ with two distinct vertices being adjacent if each coordinate is either equal or adjacent, and $X^{\boxtimes n}$ is the product of $X$ with itself $n$ times.

The value of $\Theta^*$ corresponds to the per use capacity of a channel with entanglement-assistance in the limit of many uses. We now prove an asymptotic version of the above lemma. To avoid messy notation, we define the following:
\[X^{n} := \overline{\overline{X}^{\boxtimes n}}.\]

\begin{lemma}
If $X \qarrow Y$, then $\Theta^*(\overline{X}) \le \Theta^*(\overline{Y})$.
\end{lemma}
\proof
Suppose that $X \qarrow Y$. It suffices to show that $c_0^*\left(\overline{X^n}\right) \le c_0^*\left(\overline{Y^n}\right)$ for all $n \in \mathbb{N}$. By the previous lemma this holds if $X^n \qarrow Y^n$ for all $n \in \mathbb{N}$. It is straightforward to check that two vertices $(x_1,\ldots,x_n)$ and $(x'_1,\ldots,x'_n)$ are adjacent in $X^n$ if and only if $x_i \sim x'_i$ for some $i \in [n]$.

To win the $\left(X^n, Y^n\right)$-homomorphism game Alice and Bob may play as follows: When they receive $(x_1,\ldots,x_n)$ and $(x'_1,\ldots,x'_n)$ respectively, they act as if they are playing the $(X,Y)$-homomorphism game with each coordinate to obtain $(y_1,\ldots,y_n)$ and $(y'_1,\ldots,y'_n)$, which they respond with accordingly. If the vertices they received were equal, then each coordinate was equal and therefore they responded with equal vertices. If the vertices they received were adjacent, then there exists $i \in [n]$ such that $x_i \sim x'_i$ and therefore $y_i \sim y'_i$ and thus they responded with adjacent vertices.

Therefore, $X^n \qarrow Y^n$ and so $c_0^*\left(\overline{X^n}\right) \le c_0^*\left(\overline{Y^n}\right)$ for all $n \in \mathbb{N}$.\qed

\section{Projective Rank}\label{rank}

As we have seen, the existence of a quantum homomorphism from a graph $X$ to a graph $Y$ is related to the existence of a set of projectors satisfying certain orthogonality conditions. Furthermore, when $Y$ is vertex transitive, it can be assumed that the projectors all have the same rank. Noting that the orthogonality conditions of Corollary~\ref{reformvtxtrans} exactly correspond to the adjacencies of $X \ltimes Y$, we see that a set of projectors satisfying the conditions in Corollary~\ref{reformvtxtrans} is exactly an assignment of rank $r$ projectors from $\mathbb{M}_\mathbb{C}(d,d)$ to the vertices of $X \ltimes Y$ such that adjacent vertices receive orthogonal projectors and $d = r|V(Y)|$. This motivates the following definition:

\begin{definition}
A \emph{$d/r$-projective representation} (or simply a $d/r$-representation) of a graph $X$ is an assignment of rank $r$ projectors in $\mathbb{M}_\mathbb{C}(d,d)$ to the vertices of $X$ such that adjacent vertices are assigned orthogonal projectors. We say that the \emph{value} of a $d/r$-representation is the rational number $\frac{d}{r}$.
\end{definition}

Note that a $3/1$-representation is not a $6/2$-representation, but they have the same values, namely 3.
The following corollary is the original motivation for defining projective representations, and simply amounts to a rephrasing of Corollary~\ref{reformvtxtrans}.

\begin{cor}\label{hom2rep}
If $X \ltimes Y$ has a projective representation of value $|V(Y)|$, then $X \qarrow Y$. Furthermore, if $Y$ is vertex transitive, then the converse also holds. 
\end{cor}
\proof
Suppose $X \ltimes Y$ has a projective $d/r$-representation of value $|V(Y)|$. The adjacencies of $X \ltimes Y$ exactly correspond to the orthogonalities of Condition~(\ref{cond:orthog}), and so this is satisfied by the projectors in the representation. Since $d = r|V(Y)|$, Condition~(\ref{cond:ident}) also holds. This proves the first claim.

The converse holds when $Y$ is vertex transitive since we can assume that all of the $d \times d$ projectors which give the quantum homomorphism from $X$ to $Y$ have the same rank $r$, and therefore $d = r|V(Y)|$ by Condition~(\ref{cond:ident}).\qed

Since $X \ltimes Y$ contains a clique of size $|V(Y)|$, it has a projective representation of value $|V(Y)|$ if and only if this is the minimum value of any projective representation of $X \ltimes Y$. This motivates the following definition.

\begin{definition}
The \emph{projective rank} of a graph $X$, denoted $\xi_f(X)$, is given by
\[\xi_f(X) = \inf\left\{\frac{d}{r} : X \text{ has a } d/r\text{-representation}\right\}.\]
\end{definition}
If we let $\Omega(r,d)$ be the graph whose vertices are the rank $r$ projectors in $\mathbb{M}_\mathbb{C}(d,d)$, and in which adjacency is orthogonality, then it is clear that a $d/r$-representation is simply a homomorphism to $\Omega(r,d)$. Therefore,
\[\xi_f(X) = \inf\left\{\frac{d}{r} : X \rightarrow \Omega(r,d)\right\}.\]
From this it is trivial to see that $\xi_f$ is preserved by homomorphisms.

It is not clear whether or not the infimum in the definition of $\xi_f$ is always attained for every graph, or even if it is always rational. Note that using the same trick as in the proof of Lemma~\ref{real}, we see that if a graph $X$ has a $d/r$-representation, then it has a $2d/2r$-representation using real projectors. Therefore restricting to real projectors does not change the value of $\xi_f(X)$.

\subsection{Relation to Other Parameters}\label{rankvsother}

An orthogonal representation of a graph is an assignment of nonzero vectors to its vertices such that adjacent vertices receive orthogonal vectors. The minimum dimension in which there exists an orthogonal representation of a graph $X$ is known as its \emph{orthogonal rank}, and is denoted by $\xi(X)$. This parameter has been investigated in various papers including \cite{qchrom}. Clearly, a projective representation using rank 1 projectors is equivalent to an orthogonal representation, and one can think of projective rank as a fractional version of orthogonal rank. Due to this, we have that
\[\xi_f(X) \le \xi(X)\]
for all graphs $X$.

One can also think of projective rank as a subspace version of fractional chromatic number. A homomorphism from $X$ to $K_{d:r}$ can be transformed into a $d/r$-projective representation by simply mapping an $r$-subset, $S$, of $[d]$ to the projection onto $\text{span}\{e_i \in \mathbb{C}^d : i \in S\}$, where $e_i$ is the $i^\text{th}$ standard basis vector. Since the fractional chromatic number of a graph $X$, denoted $\chi_f(X)$, is the minimum value of $d/r$ such that $X \rightarrow K_{d:r}$ \cite{AGT}, we see that
\[\xi_f(X) \le \chi_f(X)\]
for all graphs $X$.

Slightly less trivially, we have the following:
\begin{lemma}\label{xifchiq}
For any graph $X$,
\[\xi_f(X) \le \chi_q(X).\]
\end{lemma}
\proof
Suppose that $X$ has a quantum $c$-coloring, i.e.~a quantum homomorphism to $K_c$. Corollary~\ref{reformvtxtrans} implies that there exists rank $r$ projectors, $E_{xi} \in \mathbb{M}_\mathbb{C}(cr,cr)$, for all $x \in V(X)$ and $i \in [c]$ such that $E_{xi}E_{x'j} = 0$ whenever ($x = x'$ \& $i \ne j$) or ($x \sim x'$ \& $i = j$). In particular, $E_{x1}E_{x'1} = 0$ whenever $x \sim x'$ and thus the projectors $E_{x1}$ give a projective representation of $X$ of value $\frac{cr}{r} = c$. Therefore $\xi_f(X) \le c$.
\qed

Note that the above lemma is also implied by the fact that quantum homomorphisms respect projective rank, Theorem \ref{thm:xipreserved} below, since the projective rank of $K_c$ can easily be seen to be $c$.

In \cite{lovasz}, it was shown that $\bartheta(X) \le \xi(X)$ and $\bartheta(X) \le \chi_f(X)$ for any graph $X$. Here we show that $\bartheta$ also lower bounds projective rank. The proof uses the following definition of $\vartheta$:

\begin{defn}\label{LThandle}
For any graph $X$ with vertices $x_1, \ldots, x_n$, we have that
\[\vartheta(X) = \min_{\phi,c}\max_{i \in [n]} \frac{1}{(c^T \phi(x_i))^2}\]
where $\phi$ is an assignment of real unit vectors to the vertices of $X$ such that nonadjacent vertices are assigned orthogonal vectors and $c$ is any unit vector.
\end{defn}

We are now prepared to prove the above mentioned result.

\begin{theorem}\label{thetaxif}
For any graph $X$,
\[\bartheta(X) \le \xi_f(X).\]
\end{theorem}
\proof
We will show that $\vartheta(X) \le \xi_f(\overline{X})$ which is equivalent. The proof is similar to the proof of $\vartheta(X) \le \xi(\overline{X})$ given in \cite{lovasz}. Suppose that $E_x$ for $x \in V(X)$ gives a $d/r$-representation of $\overline{X}$ using only real projectors. Let
\[v_x = \frac{1}{\sqrt{r}}\vect(E_x)\]
for all $x \in V(X)$, and let
\[c = \frac{1}{\sqrt{d}}\vect(I_d)\]
Note that
\begin{align*}
v_x^T v_x &= \frac{1}{r} \vect(E_x)^T \vect(E_x) = \frac{1}{r} \tr \left(E_x E_x\right) \\
&= \frac{1}{r} \tr (E_x) = \frac{1}{r} r = 1
\end{align*}
and that
\[c^T c = \frac{1}{d} \vect(I)^T \vect(I) = \frac{1}{d} d = 1.\]
So the $v_x$ and $c$ are unit vectors. Also, if $x \not\simeq x'$, then $\tr(E_x E_{x'}) = 0$ and therefore $v_x^T v_{x'} = 0$. Therefore the vectors $v_x$ for $x \in V(X)$ give an orthonormal representation of $\overline{X}$ as required in Definition \ref{LThandle}. Furthermore,
\begin{align*}
c^T v_x &= \frac{1}{\sqrt{rd}} \vect(I)^T \vect(E_x) = \frac{1}{\sqrt{rd}} \tr(IE_x) \\
&= \frac{1}{\sqrt{rd}} \tr(E_x) = \frac{1}{\sqrt{rd}} r = \sqrt{\frac{r}{d}}
\end{align*}
for all $x \in V(X)$. This means that
\[\vartheta(X) \le \frac{1}{(\sqrt{r/d})^2} = \frac{d}{r}.\]
This shows that $\vartheta(X) \le \xi_f(\overline{X})$, and taking complements yields the theorem.\qed\\

For a vertex transitive graph, one can show that its projective rank and quantum independence number are related.

\begin{lemma}
If $X$ is vertex transitive, then $\xi_f(X) \le \frac{|V(X)|}{\alpha_q(X)}$.
\end{lemma}
\proof
Let $X$ be a vertex transitive graph and let $n = |V(X)|$, $m = \alpha_q(X)$, and let $E_{ix} \in \mathbb{M}_\mathbb{C}(d,d)$, for $i \in [m]$, $x \in V(X)$ be projectors which give a qunatum homomorphism from $K_m$ to $\overline{X}$. Since $X$ is vertex transitive, we can assume that all of the projectors $E_{ix}$ have the same rank $r$, and thus $d = rn$. For each $x \in V(X)$, let $F_x = \sum_{i \in [m]} E_{ix}$. We claim that the matrices $F_x$ give a projective representation of $X$ with value $n/m$. Since $E_{ix}E_{jx} = 0$ when $i \ne j$, the matrices $F_x$ are the sum of orthogonal projectors and are thus projectors. Furthermore, they have rank $mr$ since each $E_{ix}$ has rank $r$. If $x \sim x'$, then
\[F_x F_{x'} = \sum_{i,j \in [m]} E_{ix}E_{jx'} = 0\]
since $E_{ix}E_{jx'} = 0$ if ($i = j$ \& $x \ne x'$) OR ($i \ne j$ \& ($x \sim x'$ or $x = x'$)). Therefore, the projectors $F_x$ give a projective representation of $X$ of value $d/mr = n/m = |V(X)|/\alpha_q(X)$.\qed

Note that for vertex transitive graphs, this lemma gives an upper bound on $\alpha_q(X)$ which is strictly stronger than the bound in Corollary~\ref{omegaqupperbound}. We have that
\[\alpha_q(X) \le \frac{|V(X)|}{\xi_f(X)} \le \frac{|V(X)|}{\bartheta(X)} = \vartheta(X).\]

\subsection{Relation to Quantum Homomorphisms}

Our main result on projective rank is that, like $\bartheta$, it is respected by quantum homomorphisms.

\begin{theorem}\label{thm:xipreserved}
If $X \qarrow Y$, and $Y$ has a projective representation of value $\gamma \in \mathbb{Q}$, then $X$ has a projective representation of value $\gamma$, and therefore $\xi_f(X) \le \xi_f(Y)$.
\end{theorem}
\proof
Suppose that $X \qarrow Y$, and that $E_{xy} \in \mathbb{M}_\mathbb{C}(d',d')$ for $x \in V(X)$, $y \in V(Y)$ are the projectors which give the quantum homomorphism. Furthermore, let $F_y$ for $y \in V(Y)$ give a $d/r$-representation of $Y$. Define $F_x$ for $x \in V(X)$ as follows:
\[F_x = \sum_{y \in V(Y)} E_{xy} \otimes F_y.\]
We will show that this is a projective representation of value $\frac{d}{r}$ for $X$. First we must check that the matrices $F_x$ are projectors. Since $E_{xy}$ and $F_y$ are both projectors, their tensor product is a projector. Furthermore, if $y \ne y'$ then $E_{xy}E_{xy'} = 0$ and thus $(E_{xy} \otimes F_y)(E_{xy'} \otimes F_{y'}) = (E_{xy}E_{xy'}) \otimes (F_yF_{y'}) = 0$. Therefore, $F_x$ is the sum of mutually orthogonal projectors and is thus a projector. 
Next we must check that $F_x F_{x'} = 0$ for $x \sim x'$. We have that
\begin{align*}
F_x F_{x'} &= \left(\sum_{y \in V(Y)} E_{xy} \otimes F_y \right) \left(\sum_{y' \in V(Y)} E_{x'y'} \otimes F_{y'}\right) = \sum_{y,y' \in V(Y)} E_{xy}E_{x'y'} \otimes F_y F_{y'}.
\end{align*}
However, for $x \sim x'$, whenever $y \not\sim y'$ we have that $E_{xy} E_{x'y'} = 0$, and whenever $y \sim y'$ we have that $F_y F_{y'} = 0$. Therefore, all of the terms in the final sum above are 0 for $x \sim x'$ and thus $F_x F_{x'} = 0$ in this case.

Now all that is left to do is check that this representation has value $\frac{d}{r}$. The dimension of this representation is simply $dd'$ since the projectors $E_{xy}$ are in dimension $d'$ and the projectors $F_y$ are in dimension $d$. To compute the rank of $F_x$, we use the fact that trace is equal to rank for projectors. We have that
\begin{align*}
\tr(F_x) &= \tr\left(\sum_{y \in V(Y)} E_{xy} \otimes F_y \right) = \sum_{y \in V(Y)} \tr\left(E_{xy} \otimes F_y \right) \\
&= \sum_{y \in V(Y)} \tr\left(E_{xy}\right) \tr\left(F_y \right) = r \sum_{y \in V(Y)} \tr\left(E_{xy}\right) = r\tr(I_{d'}) = rd'.
\end{align*}

Therefore, the representation has value $\frac{dd'}{rd'} = \frac{d}{r}$ and thus $\xi_f(X) \le \frac{d}{r}$.
\qed

The contrapositive of the above theorem gives us a way to show that there is no quantum homomorphism from some graph to another, however the projective rank may be very difficult to compute (it may not be computable at all) and so this may not be very useful in practice.

\section{Concluding Remarks}

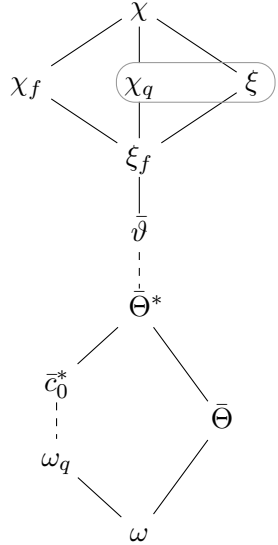
\begin{figure}[h]
\begin{center}
\begin{tikzpicture}
[vx/.style = {inner sep = 3pt, text height=1.5ex,text depth=.25ex}];

\def\hstep{-1};
\def\wstep{1.5};
\def\wstepn{1.1};

\foreach \i/\name in
         {1/$\chi$,2/$\chi_q$,3/$\xi_f$,4/$\bar{\vartheta}$,
          5/$\;\,\bar{\Theta}^*$, 8/$\omega$}{
 \node[vx](\i) at (0,\i*\hstep) {\name};}

\node[vx](2l) at (-1*\wstep,2*\hstep) {$\chi_f$};
\node[vx](2r) at (\wstep,2*\hstep) {$\xi$};
\node[vx](6l) at (-1*\wstepn,6*\hstep) {$\bar{c}_0^*$};
\node[vx](7l) at (-1*\wstepn,7*\hstep) {$\omega_q$};
\node[vx](65r) at (\wstepn,6.5*\hstep) {$\bar{\Theta}$};

\draw (1) -- (2) -- (3) -- (4);
\draw (1) -- (2l) -- (3);
\draw (1) -- (2r) -- (3);
\draw[dashed] (4) -- (5);
\draw (5) -- (6l);
\draw[dashed] (6l) -- (7l);
\draw (7l) -- (8);
\draw (5) -- (65r) -- (8);

\node[rectangle,minimum height=15pt, minimum width=60pt, 
      rounded corners=7pt,draw=black!40](rect) 
at (0.5*\wstep,2*\hstep) {};

\end{tikzpicture}
\caption{Partial Order of Graph Parameters. A circle enclosing two parameters indicates that it is not known whether they are comparable. A dashed line connecting two parameters $f$ and $g$ indicates that is not known whether $f \ne g$.}\label{Hasse}
\end{center}
\end{figure}

We have seen that quantum homomorphisms offer a natural generalization of quantum colorings and that they respect the order of some known graph parameters such as Lov\'{a}sz theta and entanglement-assisted zero-error capacity (after taking complements). We have also introduced a new parameter, projective rank, whose order is respected by quantum homomorphisms. Furthermore, we have used the notion of quantum homomorphisms to introduce quantum versions of graph parameters other than quantum chromatic number. Also, we have established various relations between these and the above mentioned parameters. As a summary, in Figure~\ref{Hasse} we give a Hasse diagram comparing the investigated graph parameters. For two graph parameters $f$ and $g$, we say that $f \le g$ if $f(X) \le g(X)$ for all graphs $X$. We also use the notation $\bar{f}(X) := f(\overline{X})$. Below we explain the relations depicted in Figure~\ref{Hasse}.

We have that $\chi \ge \chi_f,\chi_q,\xi$, since an $n$-coloring of a graph can be used to obtain fractional and quantum $n$-colorings and an orthogonal representation in dimension $n$. Furthermore, we have seen examples for which $\chi$ is strictly greater than these three parameters, such as Kneser graphs for $\chi_f$ and $\Omega_n$ for $\chi_q$ and $\xi$. Therefore $\chi > \chi_f,\chi_q,\xi$.

Since $\chi_f(C_5) = 5/2 < 3 = \chi_q(C_5) = \xi(C_5)$ and $\chi_f(\Omega_n) = |V(\Omega_n)|/\alpha(\Omega_n) > n = \chi_q(\Omega_n) = \xi(\Omega_n)$ for $n$ a sufficiently large multiple of four, we have that $\chi_f$ and $\chi_q$ are incomparable, as well as $\chi_f$ and $\xi$. The fact that $\xi_f \le \xi,\chi_f$ was discussed in Section~\ref{rankvsother}, and $\xi_f \le \chi_q$ is Lemma~\ref{xifchiq} from that section. The incomparabilities among $\chi_f, \chi_q$, and $\xi$ imply that $\xi_f$ cannot be equal to any of these, therefore $\xi_f < \chi_f, \chi_q, \xi$.

Theorem~\ref{thetaxif} states that $\bartheta \le \xi_f$, and equality does not hold since $\bartheta(C_5) = \sqrt{5}$, and it can be shown that $\xi_f(C_5) = 5/2$. In \cite{Winter} and \cite{Beigi} it is shown that $\bar{\Theta}^* \le \bartheta$, however it is an open question as to whether these two parameters are equal.

Since $\bar{\Theta}^*$ is the entanglement-assisted version of $\bar{\Theta}$, and the asymptotic version of $\bar{c}_0^*$, clearly $\bar{\Theta}^* \ge \bar{\Theta},\bar{c}_0^*$. In \cite{LMMOR} and \cite{Jop}, examples of graphs for which $\bar{c}_0^*$, and thus $\bar{\Theta}^*$, is strictly larger than $\bar{\Theta}$ are given, thus $\bar{\Theta} < \bar{\Theta}^*$. Since $\bar{c}_0^*(C_5) \le \bartheta(C_5) = \sqrt{5}$ and $\bar{c}_0^*$ is an integer, we have that $\bar{c}_0^*(C_5) = 2 < \sqrt{5} = \bar{\Theta}^*(C_5)$, and thus $\bar{c}_0^* < \bar{\Theta}^*$.

Our Theorem~\ref{thm:qindyqcap} is equivalent to $\omega_q \le \bar{c}_0^*$. However, as mentioned in Conjecture~\ref{conj:qindyqcap}, we believe that $\omega_q$ and $\bar{c}_0^*$ are in fact equal. Since an $n$-clique is also a quantum $n$-clique, we have that $\omega \le \omega_q$, and our Corollary~\ref{qindyexample} provides examples of graphs with quantum clique number strictly greater than clique number (after taking complements).

Since $\bar{\Theta}$ is the asymptotic version of $\omega$, we have that $\omega \le \bar{\Theta}$. Also, $\omega(C_5) = 2 < \sqrt{5} = \bar{\Theta}(C_5)$, so these parameters are not equal.

To see that $\bar{\Theta}$ is not comparable with either $\bar{c}_0^*$ or $\omega_q$, note that $2 = \omega(C_5) \le \omega_q(C_5) \le \bar{c}_0^*(C_5) = 2 < \sqrt{5} = \bar{\Theta}(C_5)$. Combining this with the examples from \cite{LMMOR} and \cite{Jop} mentioned above, we see that $\bar{c}_0^*$ and $\bar{\Theta}$ are incomparable. Furthermore, to obtain the lower bound on $\bar{c}_0^*$ for those examples, the authors use projective measurements and a maximally entangled state. Therefore by our Theorem~\ref{thm:qindyqcap} we have that $\omega_q$ is not comparable to $\bar{\Theta}$.\\

\subsection{Open Questions}

Perhaps the most central question is whether it is decidable if $X \qarrow Y$ for arbitrary $X$ and $Y$. If we fix the dimension of the projectors Alice and Bob are allowed to use in their strategy for the $(X,Y)$-homomorphism game, then this question reduces to determining if a particular set of quadratic equations in a finite number of variables has a solution and is therefore decidable. So bounding the dimension needed for a quantum homomorphism would resolve this problem. This question remains open even for the specific case of quantum colorings.

We have seen separations between classical and quantum versions of chromatic number and independence/clique number, and we have seen that there are no separations between the classical and quantum versions of odd girth. An interesting question is which other parameters have separations and which ones do not? More generally, for what graphs $X$ do there exist graphs $Y$ such that $X \qarrow Y$ but $X \not\to Y$? This never happens when $X$ is a cycle, but can happen for at least some complete graphs. Is there some condition on $X$ which guarantees the existence/nonexistence of such a $Y$? Similarly, for which graphs $Y$ do there exist graphs $X$ such that $X \qarrow Y$ but $X \not\to Y$? As we have seen, this happens for some complete graphs, and the only graphs for which we know this does not happen are empty graphs and bipartite graphs.

In light of Theorem~\ref{thm:qindyqcap}, it is natural to ask if $\alpha_q(X) = c_0^*(X)$ for all graphs $X$. We believe this to be the case. A positive answer would imply that the entanglement-assisted zero-error capacity of a channel can always be attained using projective measurements on a maximally entangled state, thus resolving an open problem from quantum information theory.

The definition of projective rank uses an infimum, and it is of interest whether or not that infimum can be replaced by a minimum, i.e.~that the value of $\xi_f(X)$ is always obtained by some projective representation of $X$. Even if this is not the case, one may ask if $\xi_f(X)$ is always rational. One possible approach to the former question is to try to come up with a different proof that the fractional chromatic number of a graph is always attained by some homomorphism into a Kneser graph. The current proof relies on the fact that fractional chromatic number can be written as a linear program and therefore can always be attained.

So far the only examples of graphs $X$ and $Y$ such that $X \not\rightarrow Y$ and $X \qarrow Y$ are with one of $X$ and $Y$ being complete. Note that Lemma~\ref{hom2kneser} can be used to construct examples in which neither $X$ nor $Y$ is complete. However, for the examples we constructed it turned out that $X \qarrow K_n$, $X \not\rightarrow K_n$, and $K_n$ is a subgraph of $Y$ for some $n \in \mathbb{N}$. Therefore we are interested in whether there exist graphs $X$ and $Y$ such that $X \qarrow Y$, $X \not\rightarrow Y$, and there is no $n \in \mathbb{N}$ such that $X \qarrow K_n \qarrow Y$. 

The relation ``$\rightarrow$'' gives rise to a partial order on homomorphic equivalence classes of graphs. This partial order is known as the homomorphism order of graphs, and has been heavily studied and possesses many remarkable properties \cite{Nesetril,Tardif}. In the same way, one can consider the quantum homomorphism order of graphs. The quantum homomorphism order can easily be shown to be a lattice with the same meet and join operations as the homomorphism order. Since $X \rightarrow Y \Rightarrow X \qarrow Y$, the quantum homomorphism order is a (poset) homomorphic image of the homomorphism order. It is now natural to ask which properties these orders share. In particular, is the quantum homomorphism order dense?

Although we have focused on quantum strategies for the homomorphism game, some of the proofs do not rely on the type of strategy being used. What separates results proved in this manner from results which require explicit use of the type of strategy being considered?

\paragraph{Acknowledgments.} The authors would like to thank Giannicola Scarpa for stimulating discussions and Simone Severini for his helpful notes. DR would like to thank Chris Godsil for his support throughout this project. We would also like to thank the anonymous referees of QIP 2013 for their helpful comments on our technical report.

\bibliographystyle{alphaurl}

\begin{thebibliography}{CLMW10}

\bibitem[AC85]{nohomo}
Michael~O. Albertson and Karen~L. Collins.
\newblock Homomorphisms of {$3$}-chromatic graphs.
\newblock {\em Discrete Mathematics}, 54(2):127--132, 1985.
\newblock \href {http://dx.doi.org/10.1016/0012-365X(85)90073-1}
  {\path{doi:10.1016/0012-365X(85)90073-1}}.

\bibitem[AHKS06]{Avis}
David Avis, Jun Hasegawa, Yosuke Kikuchi, and Yuuya Sasaki.
\newblock A quantum protocol to win the graph colouring game on all {H}adamard
  graphs.
\newblock {\em IEICE Transactions on Fundamentals of Electronics,
  Communications and Computer Sciences}, 89(5):1378--1381, 2006.
\newblock \href {http://arxiv.org/abs/quant-ph/0509047v4}
  {\path{arXiv:quant-ph/0509047v4}}.

\bibitem[BBG13]{Jop}
Jop Bri\"{e}t, Harry Buhrman, and Dion Gijswijt.
\newblock Violating the {S}hannon capacity of metric graphs with entanglement.
\newblock {\em Proceedings of the National Academy of Sciences},
  110(48):19227--19232, 2013.
\newblock \href {http://arxiv.org/abs/1207.1779v1} {\path{arXiv:1207.1779v1}}.

\bibitem[BCT99]{BCT99}
Gilles Brassard, Richard Cleve, and Alain Tapp.
\newblock Cost of exactly simulating quantum entanglement with classical
  communication.
\newblock {\em Physical Review Letters}, 83:1874--1877, 1999.
\newblock \href {http://arxiv.org/abs/quant-ph/9901035}
  {\path{arXiv:quant-ph/9901035}}.

\bibitem[BCW98]{BCW98}
Harry Buhrman, Richard Cleve, and Avi Wigderson.
\newblock Quantum vs. classical communication and computation.
\newblock In {\em Proceedings of the 30th Annual ACM symposium on Theory of
  Computing}, pages 63--68. ACM, 1998.
\newblock \href {http://arxiv.org/abs/quant-ph/9802040}
  {\path{arXiv:quant-ph/9802040}}.

\bibitem[Bei10]{Beigi}
Salman Beigi.
\newblock Entanglement-assisted zero-error capacity is upper-bounded by the
  {L}ov\'{a}sz theta function.
\newblock {\em Physical Review A}, 82:10303--10306, 2010.
\newblock \href {http://arxiv.org/abs/1002.2488} {\path{arXiv:1002.2488}}.

\bibitem[CHTW04]{Cleve04}
Richard Cleve, Peter Hoyer, Ben Toner, and John Watrous.
\newblock Consequences and limits of nonlocal strategies.
\newblock In {\em Proceedings of the 19th IEEE Annual Conference on
  Computational Complexity}, pages 236--249, 2004.
\newblock \href {http://arxiv.org/abs/quant-ph/0404076v2}
  {\path{arXiv:quant-ph/0404076v2}}.

\bibitem[CLMW10]{entcap}
Toby~S. Cubitt, Debbie Leung, William Matthews, and Andreas Winter.
\newblock Improving zero-error classical communication with entanglement.
\newblock {\em Physical Review Letters}, 104:230503, 2010.
\newblock \href {http://arxiv.org/abs/0911.5300} {\path{arXiv:0911.5300}}.

\bibitem[CMN{\etalchar{+}}07]{qchrom}
Peter~J. Cameron, Ashley Montanaro, Michael~W. Newman, Simone Severini, and
  Andreas Winter.
\newblock On the quantum chromatic number of a graph.
\newblock {\em Electronic Journal of Combinatorics}, 14(1), 2007.
\newblock \href {http://arxiv.org/abs/quant-ph/0608016}
  {\path{arXiv:quant-ph/0608016}}.

\bibitem[DSW11]{Winter}
Runyao Duan, Simone Severini, and Andreas Winter.
\newblock Zero-error communication via quantum channels and a quantum
  {L}ov\'asz $\vartheta$ function.
\newblock In {\em Proceedings of the 2011 IEEE International Symposium on
  Information Theory Proceedings}, pages 64--68, 2011.
\newblock \href {http://arxiv.org/abs/1002.2514} {\path{arXiv:1002.2514}}.

\bibitem[FIG11]{qchrom-aqis}
Junya Fukawa, Hiroshi Imai, and Fran\c{c}ois~Le Gall.
\newblock Quantum coloring games via symmetric {SAT} games.
\newblock In {\em Proceedings of the 11th Asian Quantum Information Science
  Conference}, 2011.

\bibitem[FR87]{frankrod}
Peter Frankl and Vojt{\v{e}}ch R{\"o}dl.
\newblock Forbidden intersections.
\newblock {\em Transactions of the American Mathematical Society},
  300(1):259--286, 1987.
\newblock \href {http://dx.doi.org/10.2307/2000598}
  {\path{doi:10.2307/2000598}}.

\bibitem[God03]{interestinggraphs}
Chris Godsil.
\newblock Interesting graphs and their colourings.
\newblock Unpublished notes, 2003.

\bibitem[GR01]{AGT}
Chris Godsil and Gordon Royle.
\newblock {\em Algebraic graph theory}, volume 207 of {\em Graduate Texts in
  Mathematics}.
\newblock Springer-Verlag, 2001.

\bibitem[GW02]{Galliard02}
Viktor Galliard and Stefan Wolf.
\newblock Pseudo-telepathy, entanglement, and graph colorings.
\newblock In {\em Proceedings of the 2002 IEEE International Symposium on
  Information Theory}, page 101, 2002.
\newblock \href {http://dx.doi.org/10.1109/ISIT.2002.1023373}
  {\path{doi:10.1109/ISIT.2002.1023373}}.

\bibitem[HN04]{Nesetril}
Pavol Hell and Jaroslav Ne\v{s}et\v{r}il.
\newblock {\em Graphs and homomorphisms}.
\newblock Oxford University Press, 2004.

\bibitem[HT97]{Tardif}
Ge\v{n}a Hahn and Claude Tardif.
\newblock Graph homomorphisms: structure and symmetry.
\newblock In {\em Graph symmetry}, volume 497, pages 107--166. Springer, 1997.
\newblock \href {http://dx.doi.org/10.1007/978-94-015-8937-6_4}
  {\path{doi:10.1007/978-94-015-8937-6_4}}.

\bibitem[KMS98]{chivec}
David Karger, Rajeev Motwani, and Madhu Sudan.
\newblock Approximate graph coloring by semidefinite programming.
\newblock {\em Journal of the ACM}, 45(2):246--265, 1998.
\newblock \href {http://arxiv.org/abs/cs/9812008} {\path{arXiv:cs/9812008}}.

\bibitem[LMM{\etalchar{+}}12]{LMMOR}
Debbie Leung, Laura Man\v{c}inska, William Matthews, Maris Ozols, and Aidan
  Roy.
\newblock Entanglement can increase asymptotic rates of zero-error classical
  communication over classical channels.
\newblock {\em Communications in Mathematical Physics}, 311:97--111, 2012.
\newblock \href {http://arxiv.org/abs/1009.1195} {\path{arXiv:1009.1195}}.

\bibitem[Lov79]{lovasz}
L{\'a}szl{\'o} Lov{\'a}sz.
\newblock On the {S}hannon capacity of a graph.
\newblock {\em IEEE Transactions on Information Theory}, 25(1):1--7, 1979.
\newblock \href {http://dx.doi.org/10.1109/TIT.1979.1055985}
  {\path{doi:10.1109/TIT.1979.1055985}}.

\bibitem[MSS13]{MSS}
Laura Man\v{c}inska, Giannicola Scarpa, and Simone Severini.
\newblock New separations in zero-error channel capacity through projective
  {Kochen-Specker} sets and quantum coloring.
\newblock {\em IEEE Transactions on Information Theory}, 59(6):4025--4032,
  2013.
\newblock \href {http://arxiv.org/abs/1207.1111} {\path{arXiv:1207.1111}}.

\bibitem[NC00]{NC}
Michael~A. Nielsen and Isaac~L. Chuang.
\newblock {\em Quantum Computation and Quantum Information}.
\newblock Cambridge University Press, 2000.

\bibitem[Rob16]{conic}
David~E. Roberson.
\newblock Conic formulations of graph homomorphisms.
\newblock {\em Journal of Algebraic Combinatorics}, 43(4):877--913, 2016.
\newblock \href {http://arxiv.org/abs/1411.6723} {\path{arXiv:1411.6723}}.

\bibitem[SS12]{SS12}
Giannicola Scarpa and Simone Severini.
\newblock {K}ochen-{S}pecker sets and the rank-1 quantum chromatic number.
\newblock {\em IEEE Transactions on Information Theory}, 58(4):2524--2529,
  2012.
\newblock \href {http://arxiv.org/abs/1106.0712} {\path{arXiv:1106.0712}}.

\bibitem[WBC15]{calculus}
Christopher~J. Wood, Jacob~D. Biamonte, and David~G. Cory.
\newblock Tensor networks and graphical calculus for open quantum systems.
\newblock {\em Quantum Information \& Computation}, 15(9\&10):759--811, 2015.
\newblock \href {http://arxiv.org/abs/1111.6950} {\path{arXiv:1111.6950}}.

\end{thebibliography}
\newcommand{\etalchar}[1]{$^{#1}$}

\end{document}